\documentclass[useAMS, onecolumn]{mn2e}

\pdfoutput=1
\usepackage{epsfig}
\usepackage{amsmath}
\usepackage{amssymb}
\usepackage{appendix}

\def\be{\begin{equation}}
\def\ee{\end{equation}}
\def\ba{\begin{eqnarray}}
\def\ea{\end{eqnarray}}
\def\go{\mathrel{\raise.3ex\hbox{$>$}\mkern-14mu
             \lower0.6ex\hbox{$\sim$}}}
\def\lo{\mathrel{\raise.3ex\hbox{$<$}\mkern-14mu
             \lower0.6ex\hbox{$\sim$}}}
\def\bxi{{\mbox{\boldmath $\xi$}}}

\def\bomega{{\bar\omega_\alpha}}

\begin{document}

\title[Tidal Heating in White Dwarf Binaries]
{Dynamical Tides in Compact White Dwarf Binaries: Helium Core White Dwarfs, Tidal Heating, and Observational Signatures}
\author[J. Fuller and D. Lai]
{Jim Fuller\thanks{Email:
derg@astro.cornell.edu; dong@astro.cornell.edu}
and Dong Lai\\
Center for Space Research, Department of Astronomy, Cornell University, Ithaca, NY 14853, USA}

\label{firstpage}
\maketitle

\begin{abstract}

Tidal dissipation in compact white dwarf (WD) binary systems significantly influences the physical conditions (such as surface
temperature and rotation rate) of the WDs prior to mass transfer or merger. In these systems, the dominant tidal effects involve the excitation of gravity waves and their dissipation in the outer envelope of the star.  We calculate the amplitude of tidally excited gravity waves in low-mass ($0.3M_\odot$) helium-core (He) WDs as a function of the tidal forcing frequency $\omega$. Like carbon-oxygen (CO) WDs studied in our previous paper, we find that the dimensionless tidal torque $F(\omega)$ (inversely proportional to the effective tidal quality factor) depends on $\omega$ in an erratic way. On average, $F(\omega)$ scales approximately as $\omega^6$, and is several orders of magnitude smaller for He WDs than for CO WDs. We find that tidal torques can begin to synchronize the WD rotation when the orbital period is less than about a hour, although a nearly constant asynchronization is maintained even at small periods.  We examine where the tidally excited gravity waves experience non-linear breaking or resonant absorption at a critical layer, allowing us to estimate the location and magnitude of tidal heating in the WD envelope. We then incorporate tidal heating in the MESA stellar evolution code, calculating the physical conditions of the WD as a function of orbital period for different WD models. We find that tidal heating makes a significant contribution to the WD luminosity for short-period ($\sim 10$~min) systems such as SDSS J0651+2844. We also find that for WDs containing a hydrogen envelope, tidal heating can trigger runaway hydrogen shell burning, leading to a nova-like event before the onset of mass transfer.

\end{abstract}


\section{Introduction}

In the last decade, compact white dwarf (WD) binaries (with orbital periods in the range of minutes to hours) have become increasingly important for several topics in astrophysics. The orbits of these systems decay via the emission of gravitational waves, constituting the largest signals for next generation space-based gravitational wave detectors. Systems of sufficiently short orbital period will merge within a Hubble time, the result of which may create a variety of exotic astrophysical systems, e.g., isolated sdB and sdO stars, R CrB stars, AM CVn binaries, or high-mass neutron stars. Of most importance, merging WDs may trigger type Ia supernovae (e.g., Webbink 1984; Iben \& Tutukov 1984). Recent observations and numerical simulations have provided some support for such ``double degenerate'' progenitors of SNe Ia.  (e.g., Gilfanov \& Bogdan 2010; Di Stefano 2010; Maoz et al.~2010; Li et al.~2011; Bloom et al.~2012; Schaefer \& Pagnotta 2012, Gonzalez Hernandez et al. 2012).

The outcome of a WD binary merger depends on the masses and compositions of the WDs and their pre-merger conditions (e.g., Segretain et al.~1997; Yoon et al.~2007; Loren-Aguilar et al.~2009; van Kerkwijk et al.~2010; Dan et al.~2012; Raskin et al.~2012). Despite the broad significance of WD mergers in astrophysics, detailed studies of the pre-merger conditions have been relatively scarce. Most studies have focused on equilibrium (non-dynamical) tides (e.g., Iben et al. 1998, Willems et al. 2010) or have parameterized the tidal effects (e.g., Piro 2011). None of these studies have sought to calculate both the magnitude and location of tidal heating, and none of them can be used to predict observational signatures of tidal heating. Such predictions are becomingly increasingly important as ongoing surveys continue to uncover new compact WD binary systems (e.g. Mullally et al. 2009; Kulkarni \& van Kerkwijk 2010; Steinfadt et al. 2010a; Kilic et al. 2012; Brown et al. 2011; see Marsh 2011 for a review).

This paper is the fourth in a series (see Fuller \& Lai 2011,2012a,2012b, hereafter Papers I, II, and III) where we systematically study dynamical tides and their observational consequences in compact WD binaries. In Paper I, we calculated the tidal excitation of discrete g-modes in carbon-oxygen (CO) WDs. We showed that the excited g-modes reach very non-linear amplitudes near the surface of the star, even far from resonance. Thus, rather than exciting discrete g-modes, the binary companion will excite a continuous train of gravity waves that propagate towards the surface of the WD, where they are likely dissipated through a combination of non-linear processes and radiative damping. 

In Paper II, we calculated the gravity wave amplitude as a function of orbital frequency using an outgoing wave boundary condition (i.e., assuming the waves damp completely in the outer layers of the WD rather than reflecting at the surface). Our calculations showed that the train of gravity waves is launched at the C-He composition gradient in the CO WD models we used. We then computed the energy and angular momentum flux carried by the waves in order to predict the orbital and spin evolution of WDs in compact systems. We found that tidal effects are negligible at large separations (orbital periods larger than about an hour), but become increasingly important at smaller periods, causing the WDs to be nearly synchronized upon merger. Furthermore, we found that the heating rate can be quite large at short orbital periods (exceeding $100 L_\odot$ just before merger, depending on the system parameters), potentially leading to an observable tidal heating signature. In Paper III, we showed that localized tidal heating in the semi-degenerate region of the hydrogen envelope can lead to thermonuclear runaways, \textquotedblleft tidal novae", potentially burning the hydrogen layer off the WD in an explosive event similar to a classical nova.

In this paper, we extend our calculations to models of low mass ($M \approx 0.3 M_\odot$) helium (He) WDs, which are relatively common among observed short-period WD binary systems (Kilic et al. 2012). Since these WDs do not have a C-He compositional gradient (although they do contain a helium-hydrogen composition gradient), the wave excitation mechanism and the resultant amplitude of the gravity waves may be quite different from their higher-mass CO counterparts. We compute the energy and angular momentum fluxes carried by the waves, and the consequent effect on the orbital and spin evolution of He WDs in compact systems.

We also attempt to calculate the observational signatures of tidal heating in both CO and He WDs. Although our calculations of the wave amplitudes are performed in the linear theory, we use these amplitudes to estimate where the waves become non-linear as they propagate toward the surface of the WD. If the waves become sufficiently non-linear, they will overturn the stratification and break, depositing their energy and angular momentum into the surface layers of the star. Also, the WD envelope can develop differential rotation due to the deposition of angular momentum carried by the gravity waves. This in turn produces a \textquotedblleft critical layer" at which wave absorption takes place due to corotation resonance. We examine the criteria for non-linear wave breaking and for wave absorption at the critical layer, and estimate where the outgoing waves deposit their energy in He and CO WDs. We then evolve WD models using the MESA stellar evolution code (Paxton et al. 2011) including a tidal heating term (which is a function of both radius and time) calculated according to various criteria. The evolution code allows us to monitor changes in the WD properties as a function of orbital period and thus allows us to make predictions of observational signatures for WD binaries as a function of their orbital period.


The paper is organized as follows. In Section 2, we calculate the amplitude of gravity waves in a He WD model as a function of the tidal forcing frequency, and compare these results to those previously obtained for the CO WD models. In Section 3, we compute the orbital and spin evolution of a He WD in a compact binary system and estimate the magnitude of the tidal heating. In Section 4, we estimate where in the WD the waves will undergo non-linear wave breaking or experience wave absorption at a critical layer, and how the location depends on the orbital period and the internal structure of the WD. In Section 5, we evolve WD models under the influence of tidal heating and predict observational signatures of the tidal heating. Finally, in Section 6, we compare our predictions to observed systems, and we discuss the uncertainties in our results and how they may be remedied by future studies.

\section{Tidal Dissipation in Helium WDs}
\label{HeWD}

\subsection{Wave Dynamics}

To calculate the amplitude of the tidally excited gravity waves in a He WD model, we use the same method described in Paper II. Here we review only the basic concepts and introduce our notations.

The Lagrangian displacement associated with the dominant (quadrupole) component of the tidal potential has the form
\be
\label{xib}
\bxi({\bf r},t) = \big[\xi_r(r) \hat{r} + \xi_\perp (r) r \nabla_\perp \big] Y_{22}(\theta,\phi) e^{-i \omega t},
\ee
where $\xi_r$ is the radial component of the displacement and $\xi_\perp$ is the perpendicular displacement, $r$ is the radius, $Y_{22}$ is the $l=m=2$ spherical harmonic, and $\omega$ is the tidal forcing frequency. In this paper, we consider the orbit of the companion to be circular and aligned such that $\omega = 2(\Omega - \Omega_s)$, where $\Omega$ is the angular orbital frequency and $\Omega_s$ is the angular spin frequency. The displacement $\bxi$ can be further decomposed into an equilibrium component $\bxi^{\rm eq}$ that describes the quasi-static ellipsoidal distortion of the star, and a dynamical component $\bxi^{\rm dyn}$ which describes the non-equilibrium wavelike response of the star to the tidal forcing. Appendix A describes the details (improving upon the treatment of Paper II) of decomposing the equilibrium and dynamical components.

We calculate the waveform of the tidally excited gravity waves by solving the linear inhomogeneous wave equations for stellar oscillations. At the center of the star, we impose the regularity boundary condition. At the outer boundary near the stellar surface, we use the (outgoing) radiative boundary condition:
\be
\label{xiperpoutbc}
\frac{d}{dr}\xi_\perp^{\rm dyn} = \Bigg[\frac{-\Big(\rho r^2/k_r)'}{2\Big(\rho r^2/k_r\Big)} - i k_r \Bigg] \xi_\perp^{\rm dyn},
\ee
where $\rho$ is the density, $k_r$ is the radial wave number,
\be
\label{kr}
k_r^2 \simeq \frac{l(l+1)(N^2 - \omega^2)}{\omega^2 r^2},
\ee
and $N$ is the Brunt-Vaisala frequency. Equation \ref{kr} is valid as long as $\omega^2 \ll L_l^2$, where 
\be
\label{L2}
L_l^2 = \frac{l(l+1)c_s^2}{r^2}
\ee
is the square of the Lamb frequency, and $c_s$ is the sound speed.

Upon solving the oscillation equations to calculate $\xi_r(r)$ and $\xi_\perp(r)$ for a given value of $\omega$, the angular momentum flux carried by the wave is
\be
\label{Jz}
\dot{J}_z(r) = 2 m \omega^2 \rho r^3 {\rm Re}\Big[ i \xi_r^{{\rm dyn}^*} \xi_\perp^{\rm dyn}\Big].
\ee
When evaluated at the outer boundary $r=r_{\rm out}$, equation (\ref{Jz}) represents the rate at which the dynamical tide deposits angular momentum into the WD envelope. The outgoing angular momentum and energy fluxes can be written as
\be
\label{Jdot}
T_{\rm tide} = \dot{J}_z(r_{\rm out}) = T_0 F(\omega),
\ee
and
\be
\label{Edot}
\dot{E}_{\rm tide} = \Omega T_0 F(\omega),
\ee
where
\be
\label{T0}
T_0 = \frac{G M'^2}{a} \bigg(\frac{R}{a}\bigg)^5,
\ee
$M'$ is the mass of the companion star, and $a$ is the orbital semi-major axis. The dimensionless function $F(\omega)$ describes the magnitude of wave excitation in the WD, and is strongly dependent on the internal structure of the WD and the tidal frequency $\omega$. In terms of the commonly used parameterization of tidal dissipation (Goldreich \& Soter 1966; Alexander 1973; Hut 1981), $F(\omega)$ is the related to the tidal phase lag and tidal $Q$ by $F(\omega)=3k_2\delta_{\rm lag}=3k_2/(2Q)$ (assuming $\Omega>\Omega_s$), where $k_2$ is the tidal Love number.

\subsection{Wave Excitation in He WDs}
\label{excite}

\begin{figure*}
\begin{centering}
\includegraphics[scale=.6]{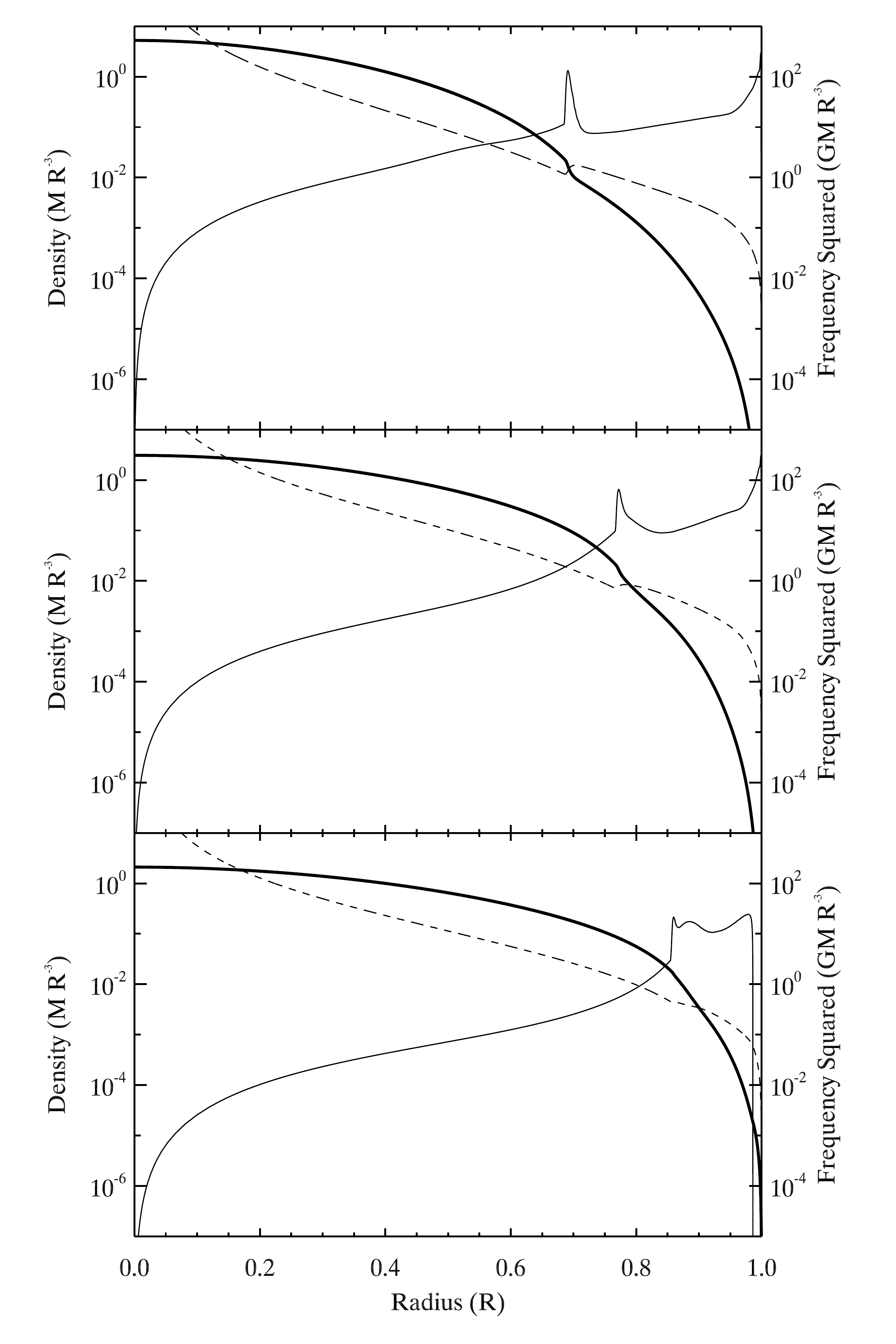}
\caption{\label{WD3struc} Propagation diagrams showing $\rho$ (thick solid line), $N^2$ (thin solid line), and $L_2^2$ (dashed line) as a function of radius in our $M=0.3 M_\odot$ He core WD model with a $\sim \! 10^{-3} M_\odot$ hydrogen shell. The three panels are for WDs with $T_{\rm eff}=18000$K (top), $T_{\rm eff}=12000$K (middle), and $T_{\rm eff}=6000$K (bottom). All quantities are plotted in units with $G=M=R=1$.}
\end{centering}
\end{figure*}

We perform our calculations on an $M=0.3 M_\odot$ He WD with a $\sim 1.2\times10^{-3} M_\odot$ hydrogen envelope, generated using the MESA stellar evolution code (Paxton et al. 2011). We evolve the same WD to surface temperatures of $T_{\rm eff} = 18000$K, $T_{\rm eff} = 12000$K, and $T_{\rm eff} = 6000$K, with respective radii of $R=2.0\times 10^9$cm, $R=1.6\times 10^9$cm, and $R=1.4\times 10^9$cm. Figure \ref{WD3struc} shows propagation diagrams for our stellar models. The spike in $N^2$ at $r\simeq 0.8 R$ is due to the He-H composition gradient.

\begin{figure*}
\begin{centering}
\includegraphics[scale=.6]{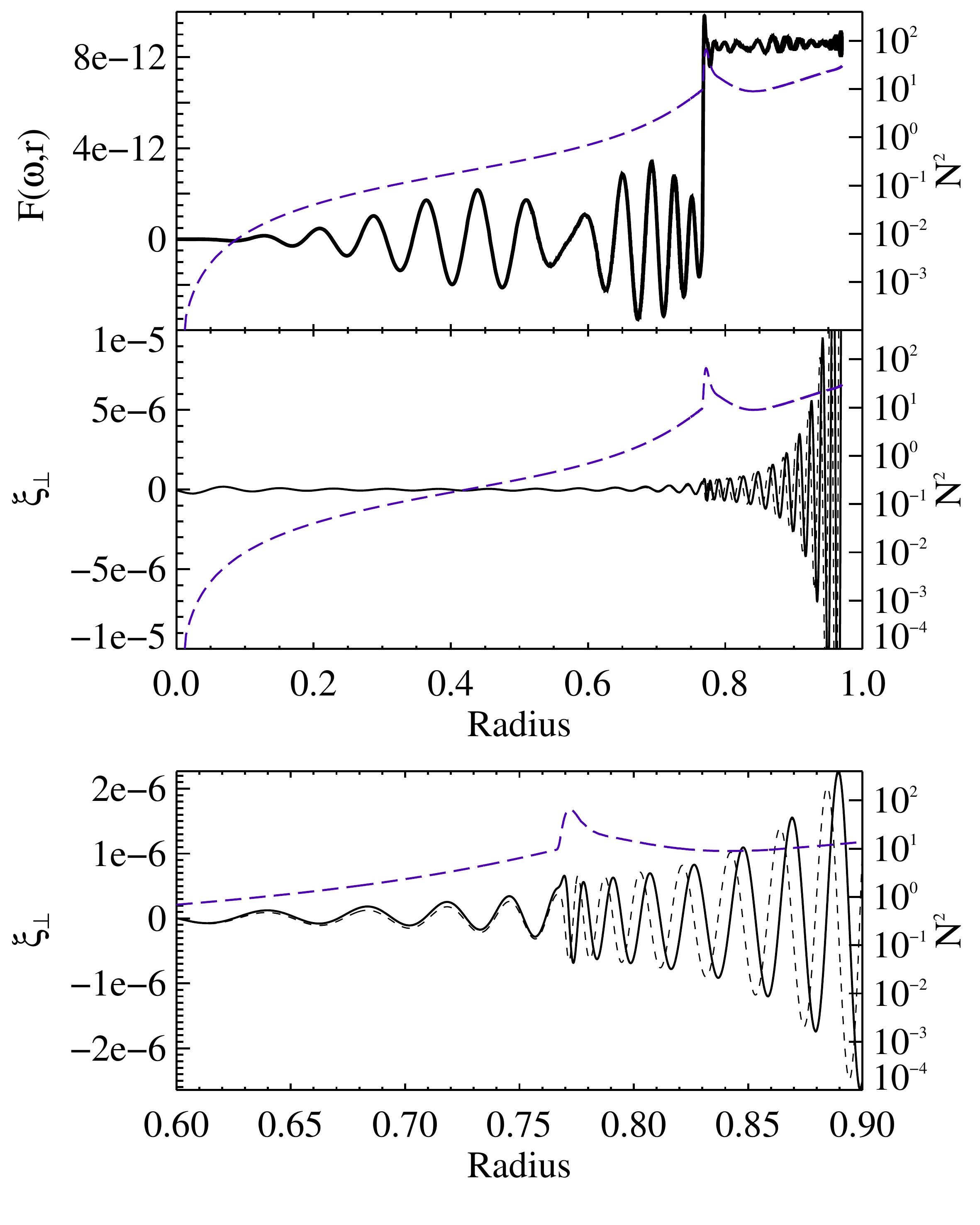}
\caption{\label{WD.3RadialEflux} Dynamical tide in our He WD model with $T_{\rm eff} = 12000$K driven by a companion of mass $M'=M$, with the tidal frequency $\omega=3.0\times10^{-2}$. Top: The value of $F(\omega,r) = \dot{J}_z(r)/T_0$ (dark solid line) as a function of radius, with $\dot{J}_z$ calculated from equation (\ref{Jz}). All values are plotted in units of $G=M=R=1$. Bottom: The real part of $\xi_\perp^{\rm dyn}$ (dark solid line) and imaginary part of $\xi_\perp^{\rm dyn}$ (dark dashed line) as a function of stellar radius. The value of $N^2$ has been plotted (dashed purple line) in each panel.}
\end{centering}
\end{figure*}

We have calculated the fluid displacement $\bxi(r)$ as a function of $r$ for tidally excited gravity waves at many values of $\omega$. Figure \ref{WD.3RadialEflux} shows the wave function $\xi_\perp(r)$ and dimensionless angular momentum flux $F(\omega,r) = \dot{J}_z(r)/T_0$ as a function of radius for $\omega = 3\times10^{-2}$, in units where $G=M=R=1$.  The location of wave excitation can be determined by examining $F(\omega,r)$. Below the wave excitation region there exists both an ingoing and outgoing wave such that $F(\omega,r) \approx 0$, while above the excitation region there exists only an outgoing wave such that $F(\omega,r) \approx $ constant. It is evident from Figure \ref{WD.3RadialEflux} that the wave is excited at the sudden increase in $N^2$ associated with the He-H composition gradient.

\begin{figure*}
\begin{centering}
\includegraphics[scale=.6]{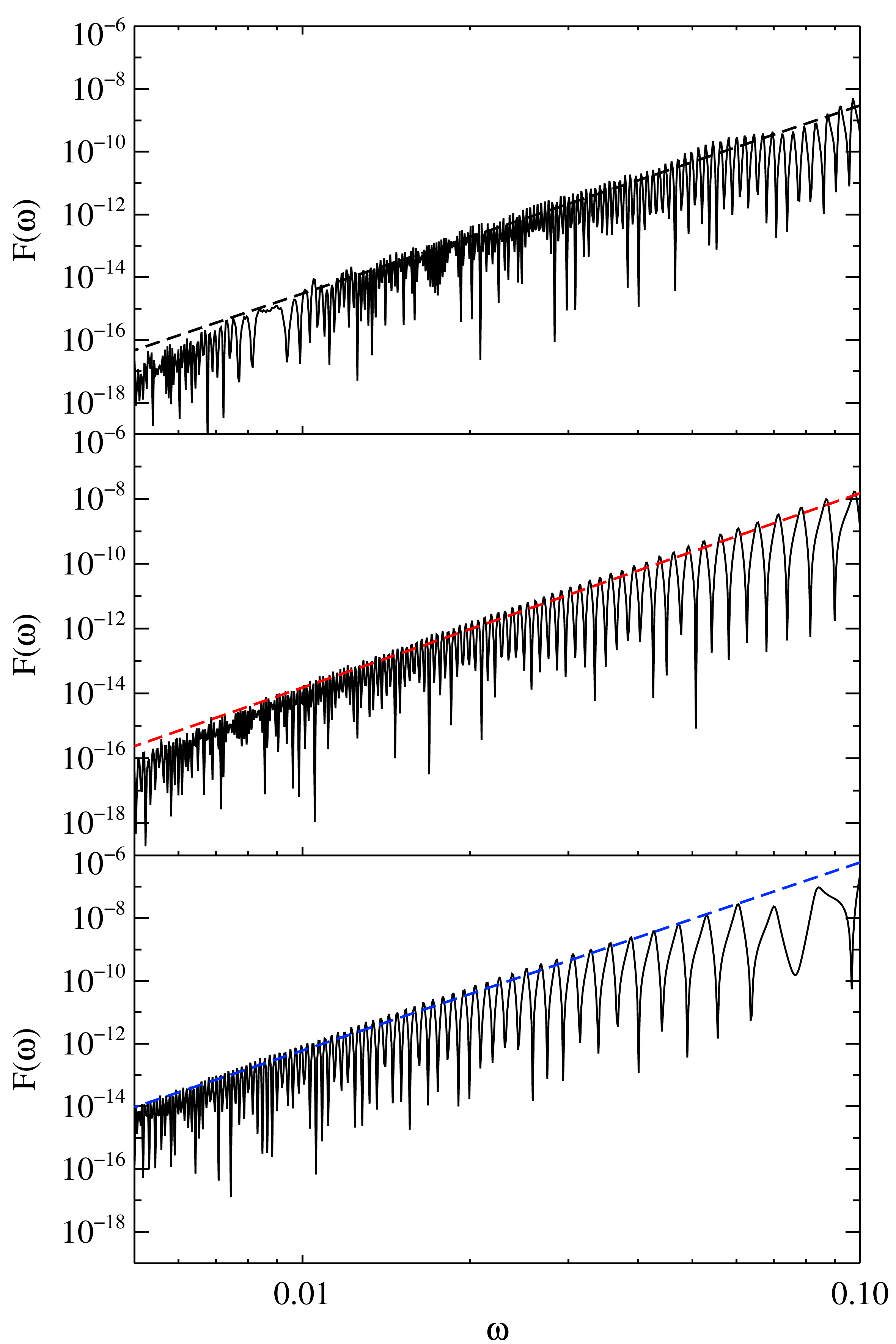}
\caption{\label{WD3F} The dimensionless tidal torque $F(\omega)= \dot{J}_z/T_o$ [see equation (\ref{Jdot})] carried by outgoing gravity waves as a function of the tidal frequency $\omega$ for our He WD model with $T_{\rm eff}=18000$K (top), $T_{\rm eff}=12000$K (middle), and $T_{\rm eff}=6000$K (bottom). The dashed lines correspond to $F(\omega)= 3\times10^{-3} \omega^6$ (top), $F(\omega)= 1.5\times10^{-2} \omega^6$ (middle), and $F(\omega)= 6\times10^{-1} \omega^6$ (bottom). The frequency is in units of $G=M=R=1$.}
\end{centering}
\end{figure*}

Figure \ref{WD3F} displays $F(\omega)$ as a function of $\omega$. It is evident that $F(\omega)$ is an erratic, non-monotonic function of $\omega$. The reason is that the core of the WD behaves as a quasi-cavity containing in-going and out-going waves. At some frequencies, these waves constructively interfere at the cavity boundary (the He-H composition gradient) and create a large outgoing wave in the WD envelope. At other frequencies, the waves exhibit deconstructive interference, creating a small outgoing wave in the envelope (see Paper II for more details). In Paper II we found that on average, $F(\omega)$ has the rough scaling $F(\omega) \propto \omega^5$ for CO WDs, whereas we find here that in a rough sense $F(\omega) \propto \omega^6$ in our He WD models. Moreover, the magnitude of $F(\omega)$ of He WDs is smaller by nearly five orders of magnitude over the frequency range of interest.

The smaller values of $F(\omega)$ in He WDs stem from the dynamics of gravity wave excitation at a composition gradient. In Paper II, we showed that gravity wave excitation due to a composition gradient in CO WDs has a rough scaling
\be
\label{Ff}
F(\omega) \approx \hat{f} \hat{\omega}^5,
\ee
where the notation $\hat{x}$ indicates that the quantity $x$ should be evaluated in dimensionless units with $G=M=R=1$, and $\hat{f}$ is approximately given by 
\begin{align}
\hat{f} &\approx \frac{\hat{\rho}_a \hat{r}_a^7 \hat{N}_a}{\hat{N}_b^4 \hat{g}_a^2}\bigg(\frac{\hat{r}_a}{\hat{H}_a}\bigg)^4 \nonumber \\
&\approx 10^{-1} \bigg(\frac{\hat{\rho}_a}{10^{-2}}\bigg) \bigg(\frac{\hat{r}_a}{0.75} \bigg)^{11} \bigg(\frac{\hat{N}_a}{8}\bigg) \bigg(\frac{\hat{N}_b}{3}\bigg)^{-4} \bigg(\frac{\hat{g}_a}{2}\bigg)^{-2} \bigg(\frac{\hat{H}_a}{0.1}\bigg)^{-4} \qquad {\rm for \ He \ WDs,} \nonumber \\
&\approx  10^3 \bigg(\frac{\hat{\rho}_a}{10^{-3}}\bigg) \bigg(\frac{\hat{r}_a}{0.8} \bigg)^{11} \bigg(\frac{\hat{N}_a}{1.5}\bigg) \bigg(\frac{\hat{N}_b}{0.3}\bigg)^{-4} \bigg(\frac{\hat{g}_a}{1.5}\bigg)^{-2} \bigg(\frac{\hat{H}_a}{0.05}\bigg)^{-4} \qquad {\rm for \ CO \ WDs,}
\label{f}
\end{align}
where $H$ is the pressure scale height. The $a$ and $b$ subscripts indicate these quantities are to be evaluated one wavelength above the base of the composition gradient, and at the base of the composition gradient, respectively. The middle line of equation (\ref{f}) is more appropriate for our He WD models, while the last line is more appropriate for our CO WD models. Waves are more easily excited at larger values of $\hat{r}_a$ due to the longer lever arm available to be torqued by the tidal potential. They are more easily excited at smaller values of $\hat{N}_b$ because smaller values in $\hat{N}_b$ produce longer wavelengths which couple better with the tidal potential.

In our He WD models, the value of $\hat{r}_a$ tends to be smaller than in CO WDs because the hydrogen layer is much thicker. Furthermore, the value of $\hat{N}_b^2$ tends to be larger in He WDs because they are less degenerate than CO WDs (i.e., they have a larger entropy and a larger entropy gradient). The difference in stellar structure coupled with the strong dependence on $\hat{r}_a$ and $\hat{N}_b$ causes the value of $\hat{f}$ to be orders of magnitude smaller in our He WD models than it is in our CO models. Furthermore, the slight dependence of $\hat{r}_a$ and $\hat{N}_a$ on $\omega$ (due to the fact that these quantities are evaluated one wavelength above the base of the composition gradient) creates a slightly steeper scaling $F(\omega) \approx \hat{f} \hat{\omega}^6$ in He WDs.

\subsection{Orbital and Rotational Evolution of He WDs}
\label{rotation}

Having computed the dimensionless tidal torque $F(\omega)$, we can calculate the rate at which energy and angular momentum are deposited in a He WD in a compact binary. In our analysis, we assume the WD maintains solid body rotation. The spin frequency of the WD evolves as
\be
\dot \Omega_s= \frac{T_0 F(\omega)}{I},
\label{Omegasdot}
\ee
where $I$ is the moment of inertia of the WD. The orbital frequency of the WD evolves due to both tidal dissipation and gravitational radiation:
\be
\dot\Omega=\frac{3T_0F(\omega)}{\mu a^2}+\frac{3\Omega}{2t_{\rm GW}}.
\label{Omegadot}
\ee
Here, $\mu$ is the reduced mass of the binary, and $t_{\rm GW} = |a/\dot{a}|$ is the gravitational wave inspiral time given by
\begin{align}
t_{\rm GW} &= \frac{5c^5}{64G^3}\frac{a^4}{MM'M_t}\nonumber\\
&= 3.2\times 10^{10} {\rm s} \bigg(\frac{M_{\odot}^2}{MM'}\bigg)\bigg(\frac{M_t}
{2M_{\odot}}\bigg)^{\!\!1/3}\!\! \bigg(\!\frac{\Omega}{0.1\,\textrm{s}^{-1}}
\bigg)^{\!\! -8/3}.
\label{tgw}
\end{align}

At large orbital periods, the second term in equation (\ref{Omegadot}) dominates the orbital and spin evolution such that $\dot{\Omega} \gg \dot{\Omega}_s$, and tidal effects are negligible. However, because of the strong dependence of $F(\omega)$ on $\omega$, tidal spin up becomes important at short orbital periods. The critical orbital frequency, $\Omega_c$, at which tidal spin up becomes important is determined by equating $\dot\Omega\simeq 3\Omega/(2t_{GW})$ and $\dot\Omega_s$.  For our He WD models, we find [cf. equations (79) and (83) of Paper II]
\begin{align}
\Omega_c&=\left[\frac{3\kappa}{10{\hat f}}\frac{M_t^{5/3}}{M'M^{2/3}}
\left(\frac{GM}{Rc^2}\right)^{5/2}\right]^{3/19}\left(\frac{GM}{R^3}\right)^{1/2}\nonumber\\
&=(7.0\times 10^{-3}{\rm s}^{-1})\,
\left(\frac{\kappa_{0.17}M_{t1}^{5/3}M_1^5
}{{\hat f}M_1' R_4^{12}}\right)^{3/19},
\label{omc}
\end{align}
where $M_t=M+M'$ and $\kappa = I/(MR^2)$. Here, $M_{t1}=(M+M')/M_\odot$, $M_1 = M/M_\odot$, $R_4 = R/(10^4{\rm km})$, and $\kappa_{0.17} = \kappa/0.17$. Our He WD models have $\kappa=0.094$, $\kappa=0.13$, and $\kappa=0.16$ for $T_{\rm eff}=18000$K, $T_{\rm eff}=12000$K, $T_{\rm eff}=6000$K, respectively.

\begin{figure*}
\begin{centering}
\includegraphics[scale=.55]{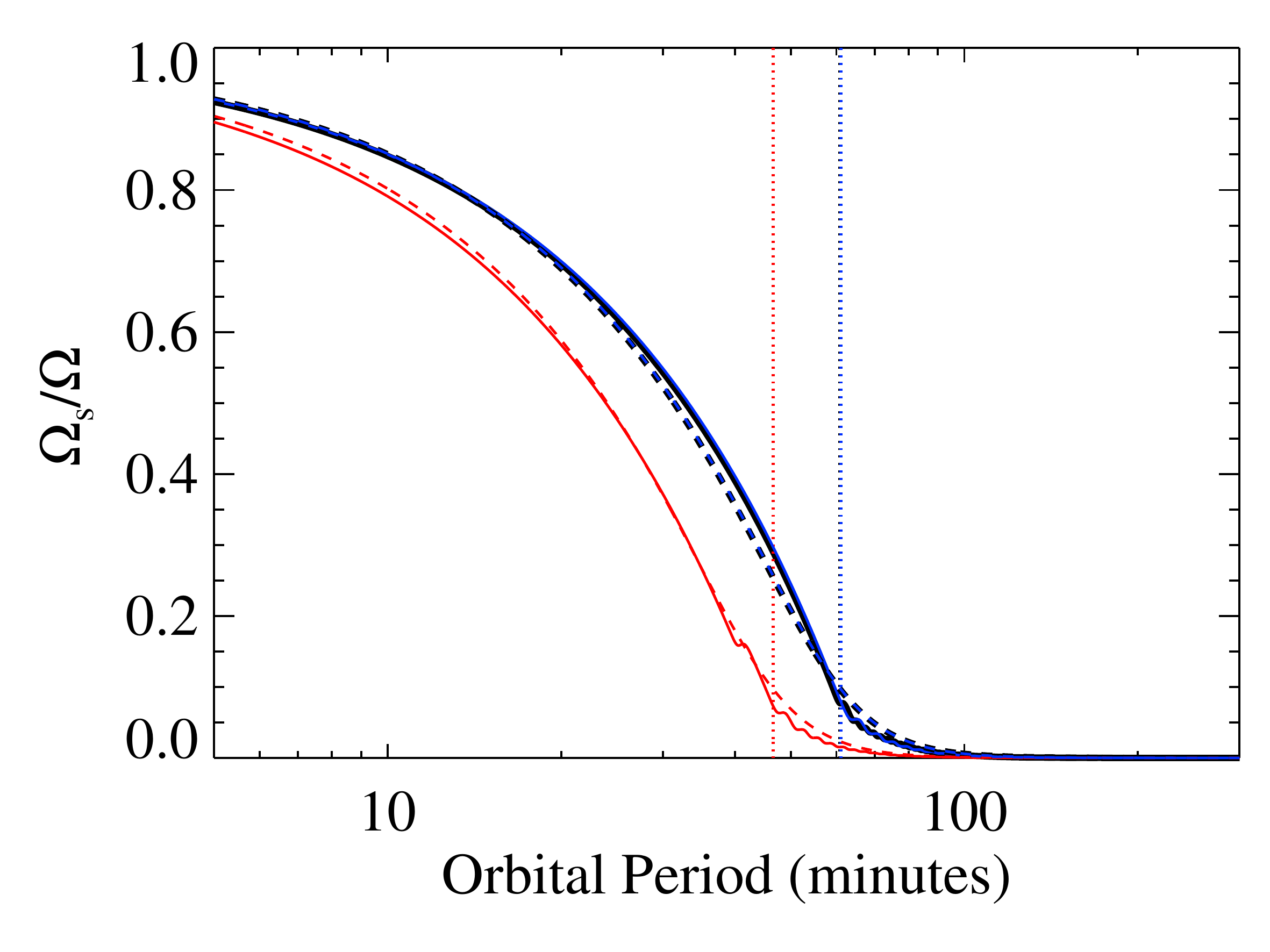}
\caption{\label{WD3SPIN} Evolution of the spin frequency $\Omega_s$ in units of the orbital frequency $\Omega$ as a function of the orbital period. The solid lines correspond to our $0.3 M_\odot$ WD model with $T_{\rm eff}=18000$K (black), $T_{\rm eff}=12000$K (red), and $T_{\rm eff}=6000$K (blue). The dashed lines correspond to evolutions using $F=3\times10^{-3}\hat{\omega}^6$ (black), $F=1.5\times10^{-2}\hat{\omega}^6$ (red), and $F=6\times10^{-1}\hat{\omega}^6$ (blue) for the WD models of like color. The vertical dotted lines denotes the critical orbital period, $2\pi/\Omega_c$ [see equation (\ref{omc})], for WDs of like color. The black and blue curves are indistinguishable from one another. In these evolutions, $M'=M$ and the WDs initially have $\Omega_s=0$.}
\end{centering}
\end{figure*}

The scaling shown in equation (\ref{omc}) is an important result of our theory.\footnote{Equations (\ref{omc}) and (\ref{Eheat2}) can be derived for the general case $T_{\rm tide} = T_0 \hat{f} \hat{\omega}^n$, as detailed in Appendix \ref{genscale}.} The dependence on $M$ and $R$ shows that lower mass, larger radius WDs (such as He WDs) should have smaller critical frequencies. So, naively, one would expect tidal dissipation to become important at longer orbital periods for low mass WDs. However, as described in Section \ref{excite}, the value of $\hat{f}$ is much smaller in He WDs. These two effects largely offset one another, causing the value of $\Omega_c$ to be similar in He and CO WDs. Thus, an He WD in a compact binary will {\it begin} to evolve toward synchronization with the orbital frequency at roughly the same orbital period as a CO WD, namely, at orbital periods $P\approx 1$ hour.

Figure \ref{WD3SPIN} displays the spin frequency, $\Omega_s$, as a function of orbital period for an He WD in a compact binary. At periods longer than about an hour, the spin frequency increases slower than the orbital frequency, and thus the value of $\Omega_s/\Omega$ remains close to zero. However, when the orbital frequency increases to $\Omega\approx\Omega_c$, the value of $\Omega_s/\Omega$ begins to increase, i.e., the WD becomes more synchronized with the orbit. The WD does not become completely synchronized, however, but retains a nearly constant degree of asynchronization such that $\Omega - \Omega_s \simeq \Omega_c$ (see Section 8.1 of Paper II for more details).

We can also calculate the amount of heat dissipated in the WD. The heat dissipated in the WD is {\it not} equal to the tidal energy transfer rate [$\dot{E}_{\rm tide}$, see equation (\ref{Edot})] because some of the tidal energy flux is stored as rotational kinetic energy. Instead, 
\be
\label{Eheat}
\dot{E}_{\rm heat} = \dot{E}_{\rm tide} (1 - \Omega_s/\Omega).
\ee
At long orbital periods where $\Omega_s/\Omega \ll 1$, $\dot{E}_{\rm heat} \simeq \dot{E}_{\rm tide}$, with $\dot{E}_{\rm tide}$ given by equation (\ref{Edot}). However, at shorter orbital periods where $\Omega > \Omega_c$ (see Paper II), 
\be
\label{etide}
\dot{E}_{\rm tide} \simeq \frac{3I\Omega^2}{2 t_{\rm GW}} \qquad {\rm for} \qquad \Omega > \Omega_c,
\ee
and we find for our He WD model that
\begin{align}
\label{Eheat2}
\dot{E}_\textrm{heat} &\simeq \dot{E}_\textrm{tide}\bigg(\frac{\Omega_c}{\Omega}\bigg)^{19/18} \nonumber \\
& \simeq (1.2\times 10^{37}\,{\rm erg~s}^{-1}) \kappa_{0.17}^{7/6}\hat f^{-1/6}M_1^{17/6}(M_1')^{5/6}(M_{t1})^{-1/18} \left(\frac{\Omega}{0.1~{\rm s}^{-1}}\right)^{65/18}. 
\end{align}
Comparison with equation (91) of Paper II reveals that the amount of heat dissipated in an He WD is the same order of magnitude as the heat dissipated in a CO WD. At short orbital periods, this heating rate can be much larger than the WD's intrinsic luminosity, and it can thus have a substantial impact on the structure, luminosity, and temperature of the WD (see Section 4).

\begin{figure*}
\begin{centering}
\includegraphics[scale=.55]{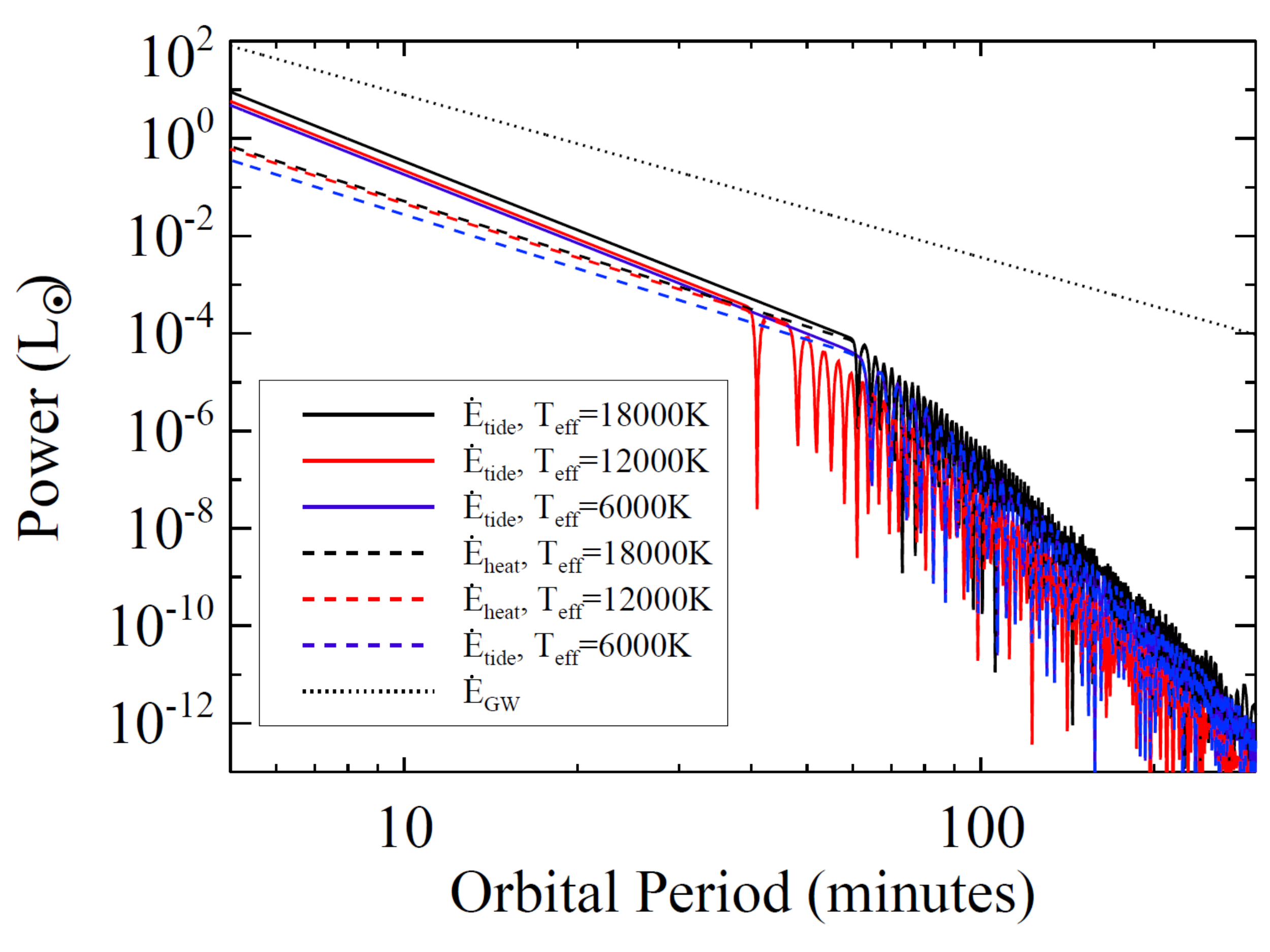}
\caption{\label{WD3ENERGY} The tidal energy dissipation rate $\dot{E}_{\rm tide}$ (solid lines) and the tidal heating rate $\dot{E}_{\rm heat}$ (dashed lines) as a function of orbital period for our $0.3 M_\odot$ WD model with $T_{\rm eff}=18000$K (black), $T_{\rm eff}=12000$K (red), and $T_{\rm eff}=6000$K (blue). Note that at long orbital periods, $\dot{E}_{\rm tide} \simeq \dot{E}_{\rm heat}$ and these curves overlap. The dotted back line is the energy flux carried away by gravitational waves, $\dot{E}_{\rm GW}$. In these evolutions, $M'=M$ and the WDs initially have $\Omega_s=0$.}
\end{centering}
\end{figure*}

Figure \ref{WD3ENERGY} shows the tidal energy flux and heating rate as a function of orbital period for a He WD in a binary with $M'=M$. It is clear that $\dot{E}_{\rm GW} \gg \dot{E}_{\rm tide}$ at all orbital periods, where $\dot{E}_{\rm GW} = GMM'/(2at_{\rm GW})$ is the energy flux carried away by gravitational waves. At long orbital periods where $\Omega < \Omega_c$, the value of $\dot{E}_{\rm heat} \simeq \dot{E}_{\rm tide}$, but the magnitude of the heating is small and has a negligible effect on the WD. At shorter orbital periods where $\Omega > \Omega_c$, the heating rate is well described by equation (\ref{Eheat2}).

\section{Location of Tidal Heat Deposition}
\label{heatlocation}

Until now, we have not examined the processes that convert the angular momentum and energy carried by gravity waves into the rotational angular momentum and internal energy of the stellar envelope. We have assumed that the gravity waves propagate into the envelope of the WD where they somehow dissipate, depositing their energy and angular momentum. Here, we investigate these processes so that we can estimate the magnitude of tidal heating as a function of depth within the star.

\subsection{Non-linear Wave Breaking}
\label{nonlinear}

As the tidally excited gravity waves propagate outward in the WD envelope, their amplitudes increase. In the WKB limit, with $\omega\ll N$ and $\omega\ll L_l$, the radial wave number is given by $k_r\simeq -k_\perp N/\omega$, where the horizontal wave number is $k_\perp=\sqrt{l(l+1)}/r$. The amplitudes of the radial and horizontal displacements scale as
\be
\label{wkb}
\xi_\perp\simeq {ik_r r\over l(l+1)}\xi_r\propto {N\over r^2 (\rho |k_r|)^{1/2}}.
\ee
Obviously, $|k_r|/k_\perp=N/\omega\gg 1$ and $|\xi_r/\xi_\perp|=\sqrt{l(l+1)} \omega/N\ll 1$.

As the gravity waves reach sufficiently large amplitudes, they are expected to break and quickly damp, locally depositing their energy and angular momentum. The critical amplitude for wave breaking may be estimated by comparing the Eulerian acceleration $\partial {\bf v}/\partial t =-\omega^2{\bxi}$ (where ${\bf v}=-i\omega {\bxi}$ is the fluid velocity) with the non-linear \textquotedblleft advective'' term ${\bf v}\cdot\nabla{\bf v} =-\omega^2 \bxi \cdot\nabla\bxi$. For gravity waves, most of the terms in the advective derivative satisfy the non-linearity condition $\bxi \cdot\nabla\bxi \simeq \bxi$ when
\be 
\label{nl1}
|k_r \xi_r| \simeq {l(l+1)\over r} |\xi_\perp|\sim 1.
\ee
Thus, nonlinear effects become important when the radial (horizontal) displacement is comparable to the radial (horizontal) wavelength. Alternatively, one may expect the waves to begin breaking when the shear is large enough to overturn the stratification of the star, i.e., when the Richardson stability criterion, $N^2/|dv_\perp/dr|^2>1/4$, is violated. This occurs when $k_\perp |\xi_\perp|\go 1$, a condition similar to equation (\ref{nl1}) for $l=2$. Equation (\ref{nl1}) is similar to the wave breaking criterion discussed in Ogilvie \& Lin (2007) and found in three-dimensional simulations of gravity waves approaching the center of a solar-type star (Barker \& Ogilvie 2011).

However, equation (\ref{nl1}) corresponds to a physical displacement, $|{\bf\xi}|\simeq |\xi_\perp|\sim R/[l(l+1)]$ in the envelope of the star. It seems unlikely that fluid displacements of order the radius of the star can be realized before non-linear wave breaking occurs. Moreover, the $\hat{r}$ component of the advective derivative contains the term $-\bxi_\perp \cdot \bxi_\perp/r$. This implies that non-linear effects may become important when $\xi_\perp^2/r \sim \xi_r$, which corresponds to $k_r \xi_\perp \sim l(l+1)$ in the WKB limit. Thus, we will also consider the non-linear breaking criterion
\be
\label{nl2}
|k_r \xi_\perp|\sim \beta,
\ee
where $\beta\go1$ is a free parameter. Choosing different values of $\beta$ will allow us to test how the location of tidal energy deposition depends on different wave breaking criteria.

To implement the non-linear criterion (\ref{nl1}) or (\ref{nl2}) for different orbital frequencies and companion masses, we can use our numerically computed wave function $\bxi(r)$ and extend it to the near surface region via the WKB amplitude relation (\ref{wkb}). Alternatively, we can use equations (\ref{Jz}), (\ref{Jdot}), and (\ref{wkb}) to find
\be
|\xi_\perp(r)| \simeq R\left({M'\over M_t}\right)\left[
{F(\omega)\over 2m\sqrt{l(l+1)}}{\Omega^4 N(r)\over G\rho(r)\omega^3}
\right]^{1/2},
\ee
with $l=m=2$. The radial displacement can be obtained from equation (\ref{wkb}). The above expression gives the wave amplitude in the outgoing wave region, i.e., the region above the composition gradient at which the waves are excited.

\begin{figure*}
\begin{centering}
\includegraphics[scale=.55]{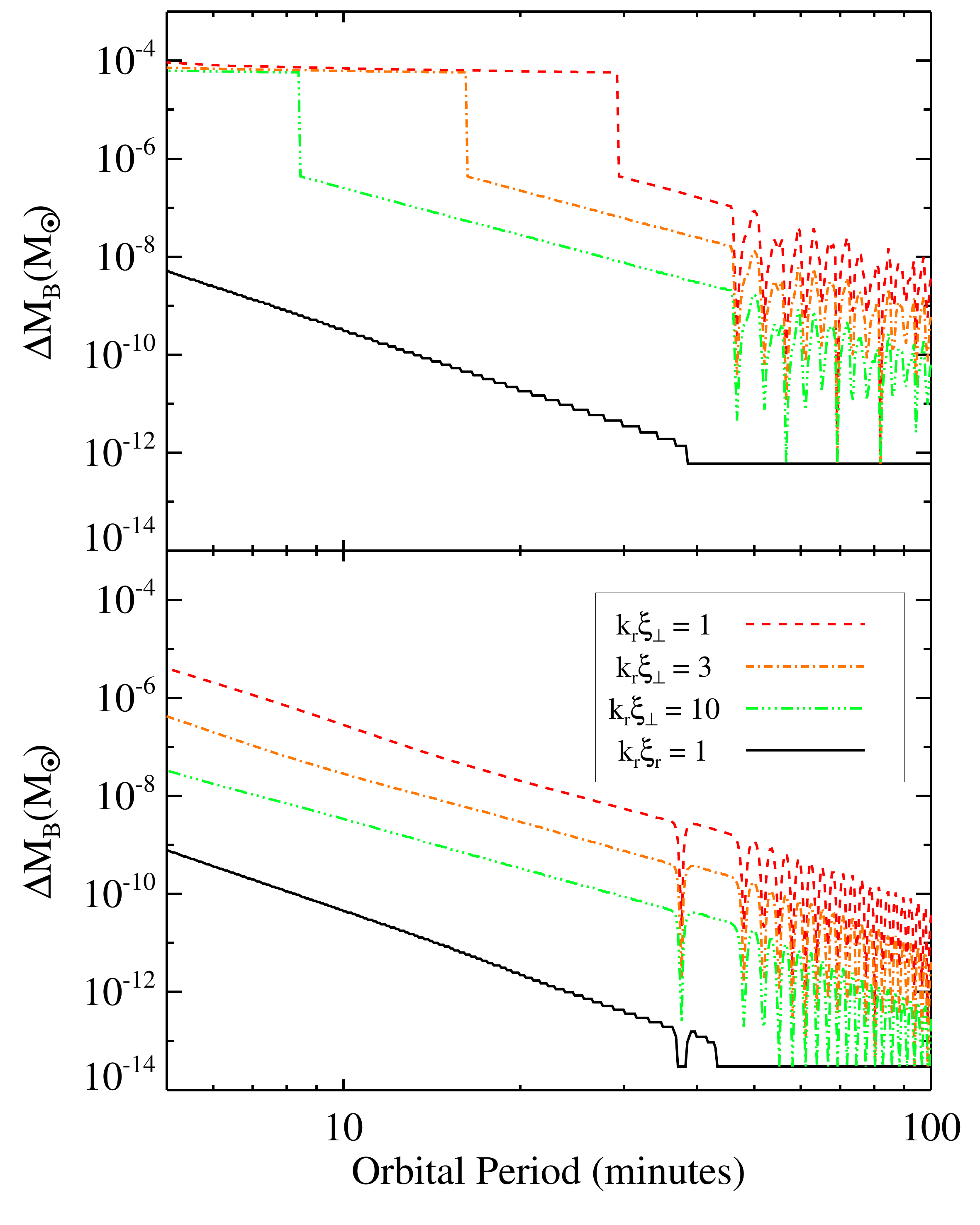}
\caption{\label{WDNL} The envelope mass $\Delta M_B$ above the point at which outgoing waves become non-linear as a function of orbital period for an $M=0.6 M_\odot$ CO WD model with $T_{\rm eff} =10000$K (top) orbiting an $M=0.3 M_\odot$ He WD model with $T_{\rm eff} =12000$K (bottom). The solid black lines are calculated from equation (\ref{nl1}), while the three dashed lines are from equation (\ref{nl2}) with $\beta=1,3,10$ (from top to bottom). This plot assumes the WD spin and orbital frequencies evolve according to equations (\ref{Omegasdot}-\ref{Omegadot}).}
\end{centering}
\end{figure*}

We denote the location of non-linear wave breaking by $\Delta M_B$, the mass above the point at which the wave amplitude satisfies one of the non-linear criteria discussed above. Figure \ref{WDNL} shows $\Delta M_B$ as a function of orbital period for both CO and He WD models, calculated according to equations (\ref{nl1}) and (\ref{nl2}). At large orbital periods ($P \gtrsim 40$ minutes), the waves do not become non-linear under criterion (\ref{nl1}) below the surface convection zone (whose depth is $\Delta M_{\rm conv} \simeq 10^{-12} M_\odot$ for the CO WD and $\Delta M_{\rm conv} \simeq 10^{-13} M_\odot$ for the He WD). In this case, the g-mode analysis of Paper I and Burkart et al. (2012) may become applicable. At shorter orbital periods, the waves become non-linear according to equation (\ref{nl1}), but they break near the surface of the WD where $\Delta M_B \lesssim 10^{-8} M_\odot$.

However, under the criterion of equation (\ref{nl2}) with $\beta=1$, the waves become non-linear deeper in the star, and nearly always reach non-linear amplitudes before encountering the surface convection zone. In the CO WD model, the value of $\Delta M_B$ jumps upward to $\Delta M_B \approx 10^{-4} M_\odot$ at an orbital period of about 30 minutes, while $\Delta M_B$ stays below $10^{-5} M_\odot$ in the He WD. The reason is that the CO WD has two composition gradients whereas the He WD has only one. The He-H composition gradient in the CO WD causes the outgoing waves excited at the C-He gradient to have larger values of both $|\xi_\perp|$ and $|k_r|$ in this layer, promoting non-linearity via equation (\ref{nl2}). No such composition gradient exists above the excitation region in an He WD, so the waves do not reach non-linear amplitudes until they are close to the surface. For reference, Figure \ref{WDNL} also shows the value of  $\Delta M_B$ calculated with the intermediate non-linearity conditions $| k_r \xi_\perp| = 3$ and $|k_r \xi_\perp| = 10$. Using these more conservative criteria yields results similar to the criterion $|k_r \xi_\perp| = 1$, but with generally smaller values of $\Delta M_B$.

\subsection{Wave Absorption at a Critical Layer}
\label{critical}

Until now, we have considered the WD to be rotating as a rigid body. Our results have indicated that in a rigidly rotating WD waves will break and deposit their angular momentum in the outer layers of the star where $\Delta M_B \lesssim 10^{-4} M_\odot$. Since a small fraction of the stellar mass absorbs the entirety of the angular momentum flux carried by waves, the outer layers of the star may spin up rapidly. If the outer layer spins up faster than angular momentum can be transported to the core, it will attain synchronous rotation with the orbit of the companion. The outgoing gravity waves will then encounter a critical layer (corotation resonance), where the wave frequency in the rest frame of the fluid, $\omega=2[\Omega-\Omega_s(r)]$, equals zero.  At the critical layer, the gravity waves are directly absorbed by the fluid since their radial wavelength $|k_r|^{-1}\simeq \omega/(k_\perp N)$ and group velocity $v_g\simeq \omega^2/(k_\perp N)$ both approach zero. As the continuous train of gravity waves deposit their angular momenta in the stellar envelope, the critical layer will move to larger depths. Such an \textquotedblleft outside-in" scenario for tidal spin-up was first discussed by Goldreich \& Nicholson (1989) for massive (early-type) stellar binaries, and was applied to WD binaries in Paper III.

In this section, we model the spin evolution of the WD using a simple two-zone model. In this model, the envelope of the star forms a critical layer that rotates synchronously with the orbit ($\Omega_{s,{\rm env}} = \Omega$), while the core of the star rotates sub-synchronously ($\Omega_{s, {\rm core}} < \Omega$). The envelope and core are coupled, with angular momentum being transferred to the core according to a parameterized coupling time, $t_{\rm coup}$. The angular momentum of the system evolves according to
\be
\label{Jedot}
\dot{J}_{\rm env} = \dot{J}_z(\omega_{\rm core}) - \frac{I_{\rm env}}{t_{\rm coup}} (\Omega_{s,{\rm env}} - \Omega_{s,{\rm core}})
\ee
\be
\label{Jcdot}
\dot{J}_{\rm core} = \frac{I_{\rm env}}{t_{\rm coup}} (\Omega_{s,{\rm env}} - \Omega_{s,{\rm core}}).
\ee
Here, $\dot{J}_z$ is the angular momentum flux which can be calculated from equation (\ref{Jdot}). We assume that the waves are excited in the core and dissipated in the envelope. Consequently, the angular momentum source term $\dot{J}_z$ is only present in the envelope evolution equation, although it is dependent on the tidal frequency in the core, $\omega_{\rm core} = 2(\Omega - \Omega_{s,{\rm core}})$. Using $\Omega_{s,{\rm env}} = \Omega$ and the fact that the star's total moment of inertia is $I=I_{\rm env}+I_{\rm core}$, equations (\ref{Jedot}) and (\ref{Jcdot}) may be rewritten 
\be
\label{Iedot}
\dot{I}_{\rm env} = - \frac{\dot{\Omega}}{\Omega}I_{\rm env} + \frac{\dot{J}_z(\omega_{\rm core})}{\Omega} - \frac{\Omega - \Omega_{s,{\rm core}}}{\Omega t_{\rm coup}} I_{\rm env}
\ee
\be
\label{Omscdot}
\dot{\Omega}_{s,{\rm core}} = \frac{\dot{I}_{\rm env}}{I-I_{\rm env}} \Omega_{s,{\rm core}} + \frac{I_{\rm env}}{I-I_{\rm env}} \frac{\Omega - \Omega_{s,{\rm core}}}{t_{\rm coup}}.
\ee
Since the orbital decay is dominated by the emission of gravitational waves, $\dot{\Omega}/\Omega \simeq 3\Omega/(2t_{\rm GW})$. 

With appropriate initial conditions, we can integrate equations (\ref{Iedot}) and (\ref{Omscdot}) to calculate the values of $I_{\rm env}$ and $\Omega_{s,{\rm core}}$ as a function of orbital period. We then obtain the mass $\Delta M_{\rm env}$ of the envelope corresponding to the value of $I_{\rm env}$. In this simple two-zone model, the tidal heat is deposited entirely at the base of the envelope where $\Delta M = \Delta M_{\rm env}$. The thickness of the envelope is dependent on the parameter $t_{\rm coup}$. Unfortunately, angular momentum transport in stars is not well understood. In stably stratified stars like WDs, angular momentum is likely transported by magnetic torques, e.g., via the Tayler-Spruit dynamo (Spruit 2002). In Appendix \ref{tcoup} we estimate value of $t_{\rm coup}$, and find that $1 {\rm yr} \lesssim t_{\rm coup} \lesssim 10^{4} {\rm yr}$ for realistic WD parameters.

\begin{figure*}
\begin{centering}
\includegraphics[scale=.55]{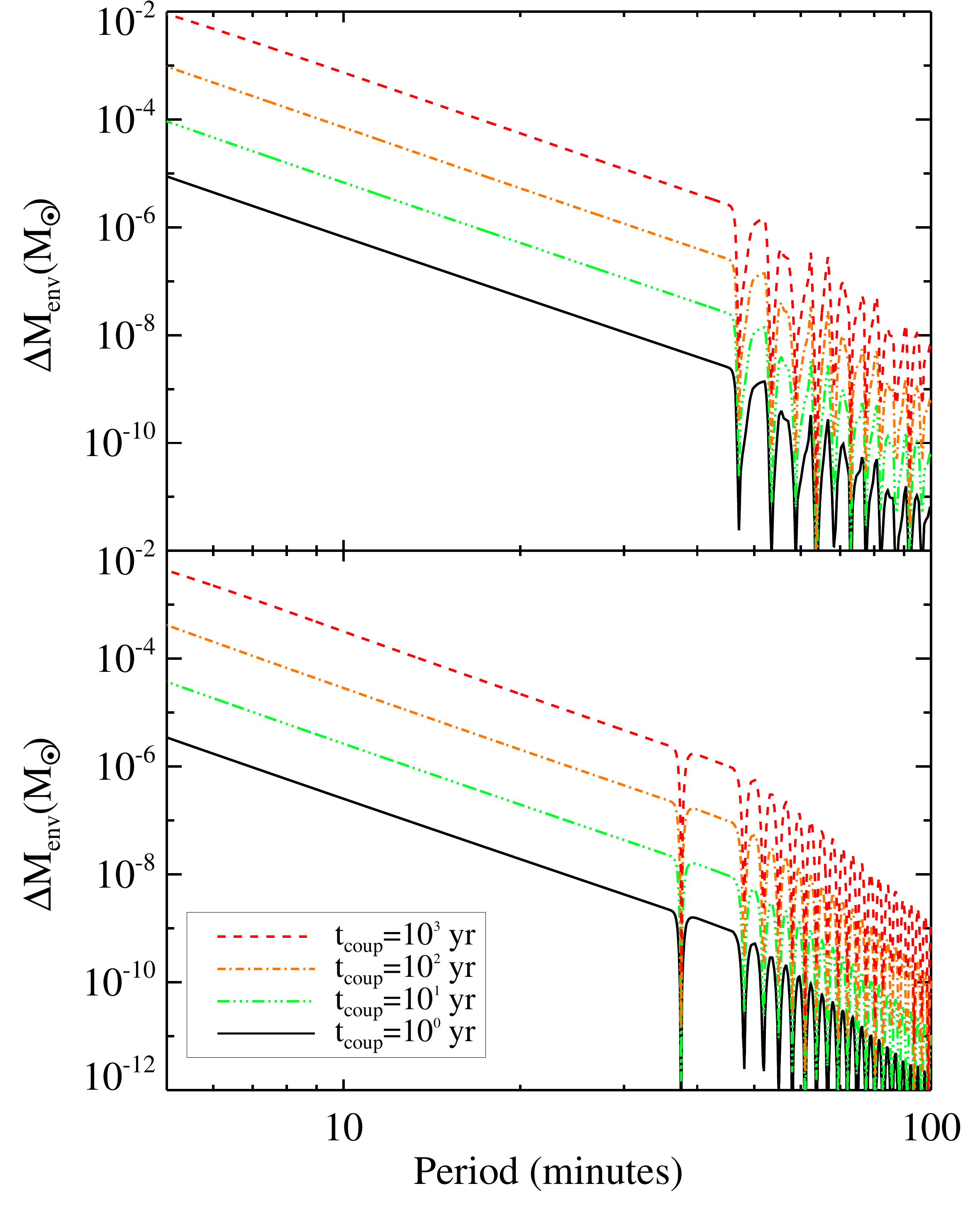}
\caption{\label{WDcrit} The envelope mass $\Delta M_{\rm env}$ above the critical layer as a function of orbital period for our $M=0.6 M_\odot$ CO WD model with $T_{\rm eff} =10000$K (top) orbiting our $M=0.3 M_\odot$ He WD model with $T_{\rm eff} =12000$K (bottom). The solid black lines have $t_{\rm coup} = 1 {\rm yr}$, the dot-dot-dashed green lines have $t_{\rm coup} = 10 {\rm yr}$, the dot-dashed orange lines have $t_{\rm coup} = 10^2 {\rm yr}$, and the dashed red lines have $t_{\rm coup} = 10^3 {\rm yr}$.}
\end{centering}
\end{figure*}

We can calculate approximate equilibrium solutions to equations (\ref{Iedot}) and (\ref{Omscdot}). Since $t_{\rm coup} \ll t_{\rm GW}$, the first term on the right hand side of equation (\ref{Iedot}) is negligible. Then at equilibrium, 
\be
\label{Ieeq}
I_{\rm env} \simeq \frac{\dot{J}_z(\omega_{\rm core}) t_{\rm coup}}{\Omega-\Omega_{s,{\rm core}}}.
\ee
At large orbital periods where $\Omega \lesssim \Omega_c$ [see equation (\ref{omc})] the system is not synchronized such that $\Omega_{s,{\rm core}} \ll \Omega$ and 
\be
\label{Ieeq1}
I_{\rm env} \simeq \frac{\dot{J}_z(2\Omega) t_{\rm coup}}{\Omega} \qquad {\rm for} \quad \Omega < \Omega_c
\ee
At short orbital periods ($\Omega > \Omega_c$) where synchronization has begun, $\Omega - \Omega_{s,{\rm core}} \simeq \Omega_c$. In this regime, $\dot J_z=\dot E_{\rm tide}/\Omega\simeq 3I\Omega/(2t_{\rm GW})$, and we have
\be
\label{Ieeq2}
I_{\rm env} \simeq \frac{3 \Omega t_{\rm coup}}{2 \Omega_c t_{\rm GW}} I \qquad {\rm for} \quad \Omega > \Omega_c.
\ee
Comparison with our numerical integration of equations (\ref{Iedot}) and (\ref{Omscdot}) shows that the approximations of equations (\ref{Ieeq1}) and (\ref{Ieeq2}) are very accurate.

Figure \ref{WDcrit} plots the value of $\Delta M_{\rm env}$ as a function of orbital period for our two WD models, using values of $t_{\rm coup}$ ranging from $1 {\rm yr}$ to $10^3 {\rm yr}$. We begin our calculation at $P_{\rm orb} > 1 {\rm hr}$ and use $I_{\rm env,0}=0$ and $\Omega_{s,{\rm core}}=0$, as is appropriate at long orbital periods where tidal effects are small. For the chosen values of $t_{\rm coup}$, the value of $\Delta M_{\rm env}$ remains small at all orbital periods ($\Delta M_{\rm env} \lesssim 10^{-2} M_{\odot}$). Thus, if a critical layer develops in a real WD, we expect it to be restricted to the outer region of the star. However, the critical layer extends to very large optical depths, suggesting that binary WDs may be observed to be synchronized at large orbital periods even if their cores are not synchronized. Our results indicate that, for our CO WD model, the critical layer most likely does not penetrate as deeply as the C-He composition gradient where gravity waves are excited (or the He-H gradient in our $0.3 M_\odot$ He WD model), so our assumption that $\dot{J}_z$ is a function of $\omega_{\rm core}=2(\Omega-\Omega_{s,{\rm core}})$ is justified. Finally, we note that the values of $\Delta M_{\rm env}$ are similar in magnitude to the values of $\Delta M_B$ calculated in Section \ref{nonlinear}.

\section{Thermal Evolution of Tidally Heated Binary White Dwarfs}
\label{heat}

In Section \ref{HeWD} (and in Paper II), we demonstrated that the tidal heating rate $\dot{E}_{\rm heat}$ of a WD in a compact binary may be substantially larger than the intrinsic luminosity of the WD. However, the consequence of this tidal heating is not clear a priori and depends on the location of heat deposition (Section \ref{heatlocation}). The assumption that the WD reaches a state of thermal equilibrium is not justified because the WD cooling time can be much larger than the gravitational inspiral time. In this section, we calculate the effect of the tidal heating on the WD structure and on its surface temperature and luminosity.

\subsection{Details of Calculation}

To calculate the effect of tidal heating, we first calculate the value of $\Delta M_B$ according to the wave breaking criteria of Sections \ref{nonlinear} and \ref{critical} (for the purposes of this calculation, we refer to the value of $\Delta M_{\rm env}$ as $\Delta M_B$, because it determines the depth at which tidal heat is deposited). We deposit the tidal heat uniformly per unit mass in the outer layers of the WD that have $\Delta M < \Delta M_B$. Although the radial dependence of this heating function is unlikely to be realistic, we find that the results are not strongly dependent on the form of the radial heat deposition (although they are sensitive to the value of $\Delta M_B$). The heating rate per gram of material, $\dot{\varepsilon}_{\rm heat}$, is then
\begin{align}
\label{epsheat1}
&\dot{\varepsilon}_{\rm heat} = 0 \quad {\rm for} \quad \Delta M > \Delta M_B \\
\label{epsheat2}
&\dot{\varepsilon}_{\rm heat} = \frac{\dot{E}_{\rm heat}}{\Delta M_B} \quad {\rm for} \quad \Delta M < \Delta M_B,
\end{align}
with the value of $\dot{E}_{\rm heat}$ calculated from equation (\ref{Eheat2}) (and its counterpart in Paper II for CO WDs). 

To understand the effect of tidal heating on the WD properties, we evolve WD models using the extra heating term calculated via equation (\ref{epsheat2}). We use the one-dimensional stellar evolution code MESA (Paxton et al. 2011) to evolve our WD models. At the beginning of the evolutions, the WDs have the same profiles used to calculate the magnitude of wave excitation (e.g., the models shown in Figure \ref{WD3struc} for our He WD models). The initial orbital period is one hour, with companion masses discussed below. 

During the course of our evolutions, we do not calculate new values of $F(\omega)$ and $\Omega_c$ at each time step. In principle, these values change as the stellar structure adjusts to tidal heating (and WD cooling). However, since waves are excited at a depth well below where they deposit their energy as heat, the properties of the WD at the location of wave excitation experience little change during the evolution. Hence, we expect the values of $F(\omega)$ and $\Omega_c$ to remain roughly constant. In contrast, the value of $\Delta M_B$ is strongly dependent on orbital period, and is updated at each time step. 

In general, the amount of tidal heat depends on the masses of the two WDs. We do not attempt to cover the whole spectrum of WD masses. Instead, we consider only two cases: a $0.6 M_\odot$ CO WD paired with a $0.3 M_\odot$ He WD (in this case we examine the heating of both WDs), and a $0.6 M_\odot$ CO WD paired with a $0.9 M_\odot$ companion star (in this case we examine only the $0.6 M_\odot$ WD). We present results for the $0.6 M_\odot$ CO WDs at initial temperatures of $5000$K, $10000$K, and $15000$K, and for the $0.3 M_\odot$ He WDs at initial temperatures of $6000$K, $12000$K, and $18000$K. These temperatures roughly span the observed temperatures in compact WD binaries (Kilic et al. 2012).

\subsection{Effects of Tidal Heating}

\begin{figure*}
\begin{centering}
\includegraphics[scale=.55]{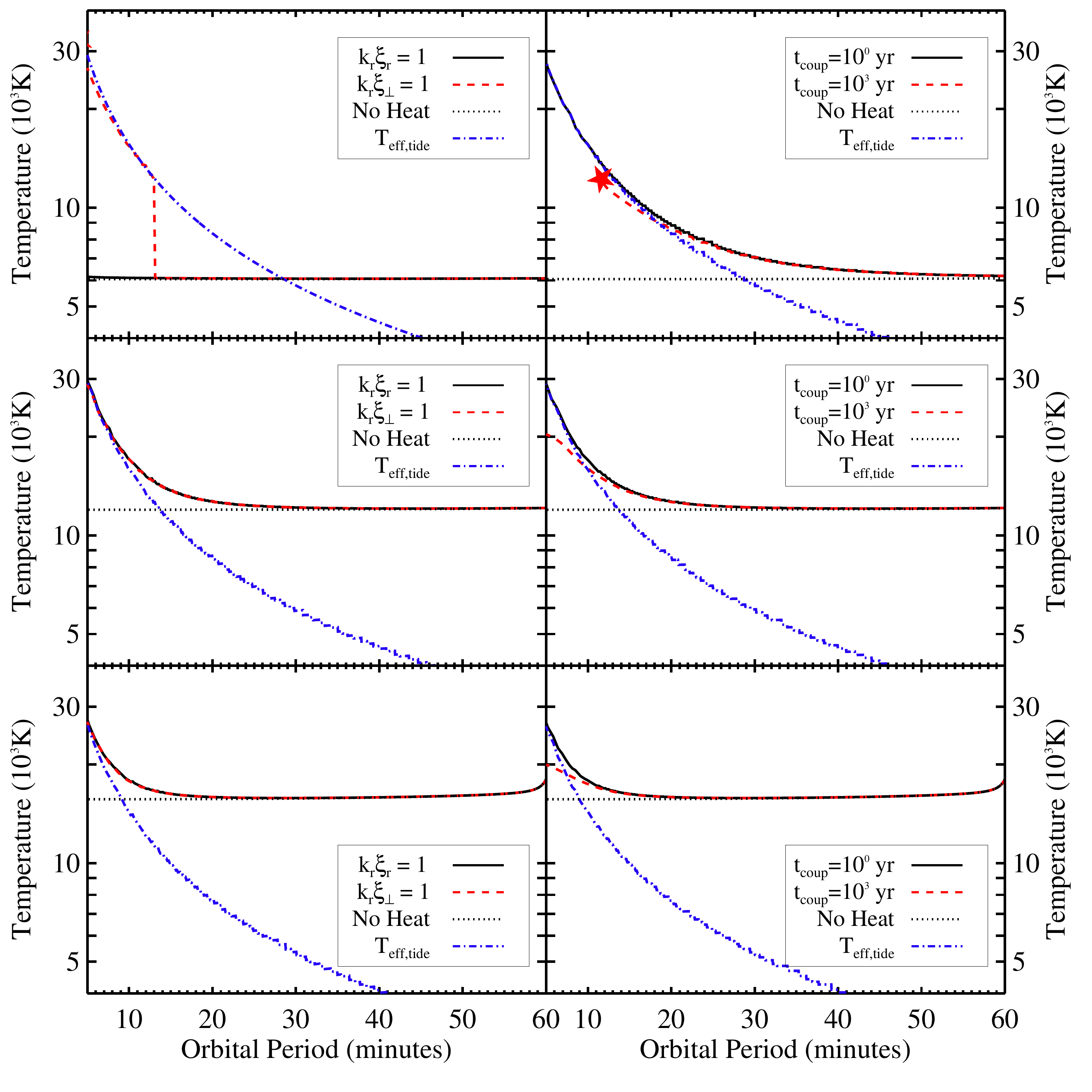}
\caption{\label{36temp} The surface temperature of our $0.3 M_\odot$ He WD model with a $0.6 M_\odot$ companion as a function of orbital period, for initial temperatures of $6000$K (top), $12000$K (middle), and $18000$K (bottom). On the left-hand side, the solid black lines are calculated with the wave breaking criterion of equation (\ref{nl1}), while the dashed red lines are calculated with equation (\ref{nl2}). On the right-hand side, the solid black lines are calculated via the two-zone model with $t_{\rm coup} = 1$yr, while the dashed red lines are calculated with $t_{\rm coup} = 10^3$yr. The dotted lines are calculated for a WD with no tidal heating and the same initial temperature. The blue dot-dashed lines are the values of and $T_{\rm eff,tide}$ from equation (\ref{tidaltemp}). The black and red lines overlap at large orbital periods, while the black and blue lines often overlap at small orbital periods. Discontinuities in temperature are due to sudden changes in the location of heat deposition (see text). Stars indicate the occurrence of a tidal nova. The plot extends to an orbital period of 5 minutes, the approximate orbital period at which Roche lobe overflow occurs.}
\end{centering}
\end{figure*}

\begin{figure*}
\begin{centering}
\includegraphics[scale=.55]{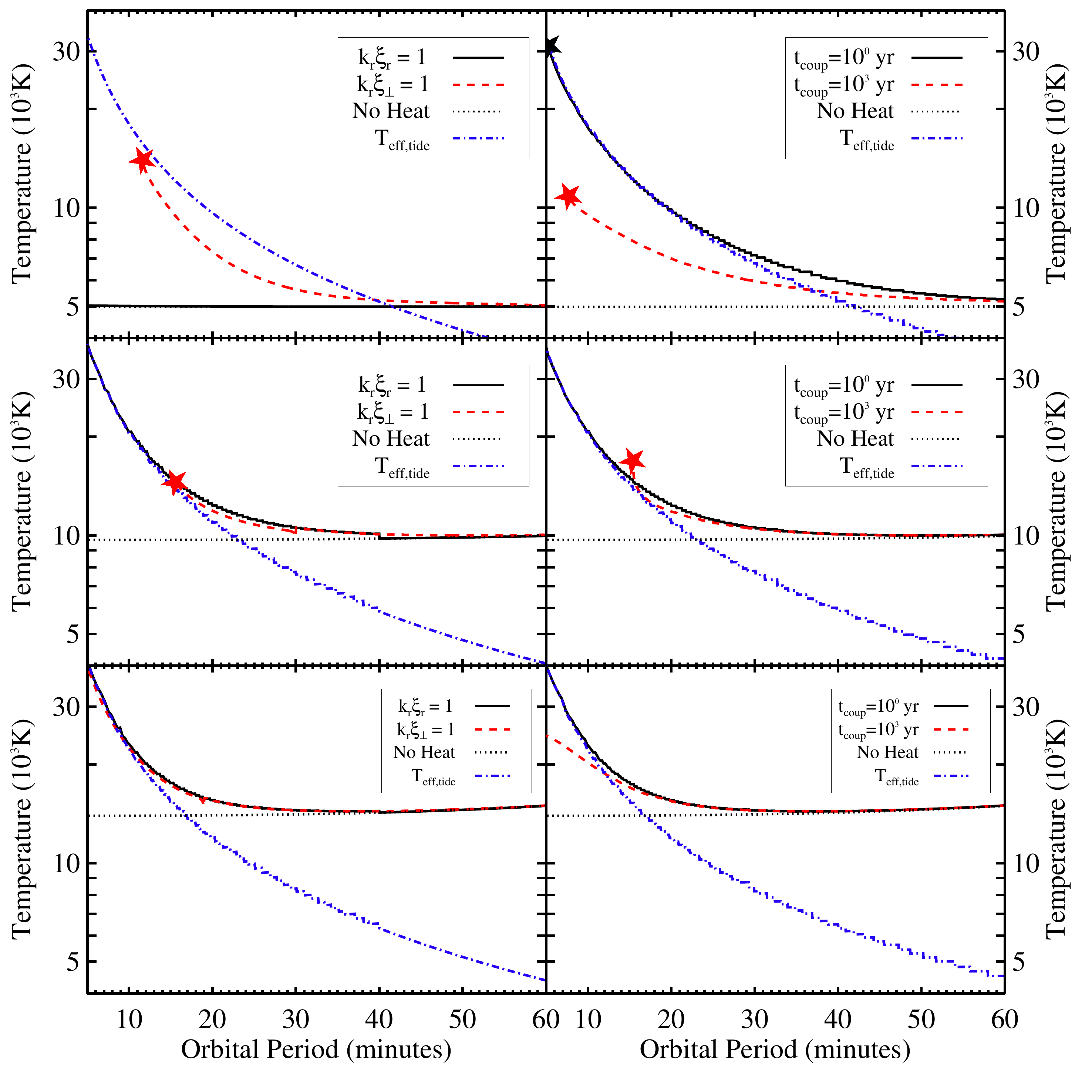}
\caption{\label{63temp} Same as Figure \ref{36temp}, but for our $0.6 M_\odot$ CO WD model with a $0.3 M_\odot$ companion.}
\end{centering}
\end{figure*}

\begin{figure*}
\begin{centering}
\includegraphics[scale=.55]{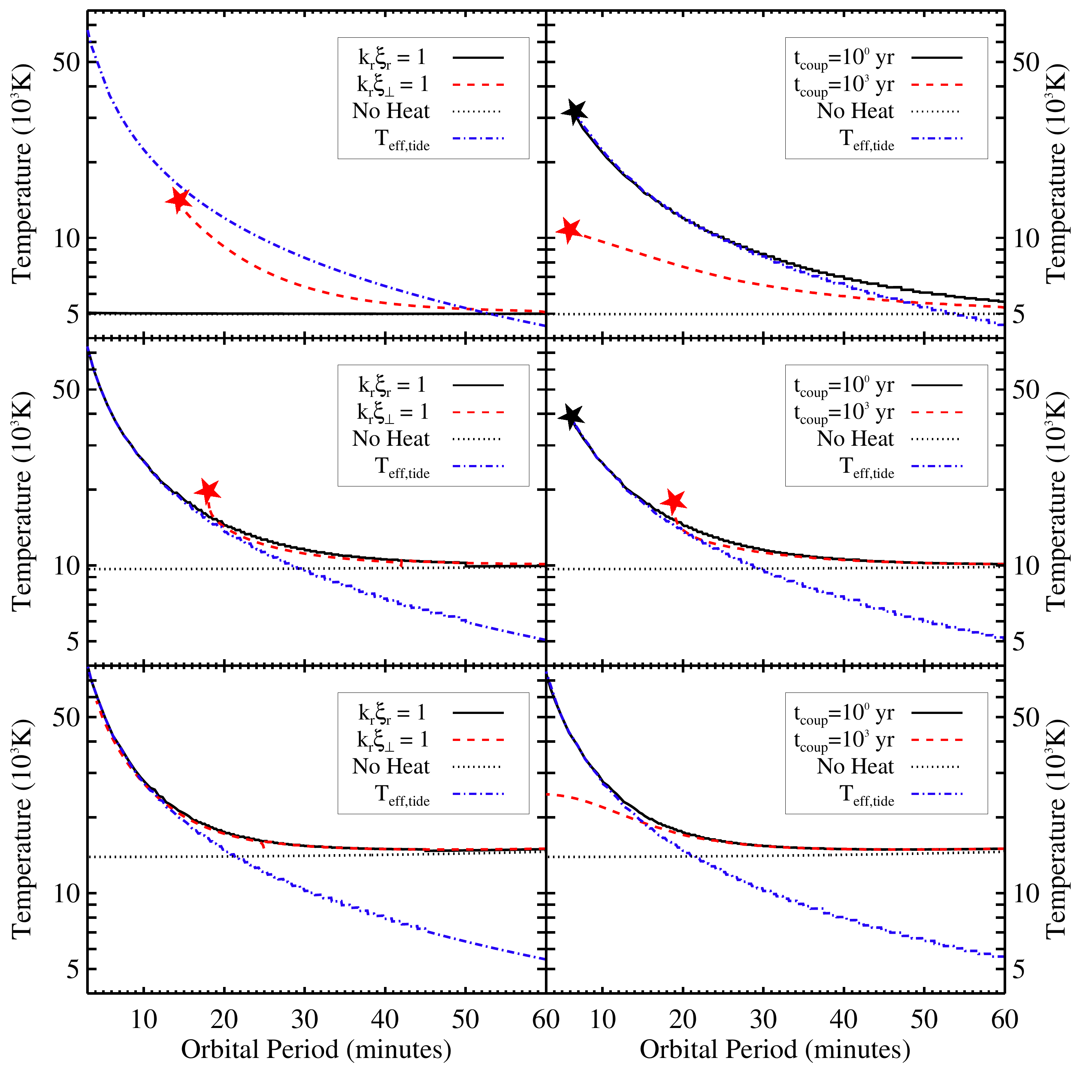}
\caption{\label{69temp} Same as Figure \ref{36temp}, but for our $0.6 M_\odot$ CO WD model with a $0.9 M_\odot$ companion.}
\end{centering}
\end{figure*}

Figure \ref{36temp} displays the surface temperature of the $0.3 M_\odot$ He WD as a function of orbital period. For comparison, we also plot the temperature of a non-tidally heated WD (i.e., we set $\dot{E}_{\rm heat}=0$ in our evolutions). At long orbital periods ($P \gtrsim 30$ minutes), the tidal heating has little effect on the luminosity or temperature of the WD, except to slightly raise the temperature of the $T_0=5000$K model. In this phase of its evolution, the WD remains well described by the cooling track for a non-tidally heated WD. At shorter periods ($P \lesssim 20$ minutes), the luminosity and temperature become substantially larger due to tidal heating. At the smallest orbital periods ($P \lesssim 8$ minutes), the WD has a temperature in excess of $18000$K regardless of its initial temperature, and the luminosity is dominated by escaping tidal heat. The case with $t_{\rm coup}=10^3$yr and initial temperature $T_{\rm eff}=6000$K (top right panel) ends abruptly, indicating that the surface hydrogen layer ignited to create a tidally induced nova, which we discuss in greater detail below (see also Paper III).

The top left panel of Figures \ref{36temp}, \ref{63temp} and \ref{69temp} are calculated for cool WDs with deep convection zones. In these WDs, the waves do not become non-linear below the convection zone according to equation (\ref{nl1}) at any orbital period, and they may become non-linear according to equation (\ref{nl2}) at orbital periods below one hour. When the waves do not become non-linear, we distribute tidal heat evenly throughout the WD, but a more accurate calculation should use the discrete g-mode formalism discussed in Paper I (c.f. Burkart et al. 2012, Valsecchi et al. 2012).

Figures \ref{36temp}-\ref{69temp} also show the tidal heating temperature, defined as
\be
\label{tidaltemp}
T_{\rm eff,tide}= \bigg(\frac{\dot{E}_{\rm heat}}{4\pi R^2 \sigma}\bigg)^{1/4}, 
\ee
where $\sigma$ is the Stefan-Boltzmann constant. In the $0.3 M_\odot$ WD, the tidal heat diffuses to the surface very quickly (see Section \ref{time} for more discussion), regardless of the non-linear breaking criterion. In this case, the observed luminosity is roughly $L \simeq L_{\rm int} + \dot{E}_{\rm heat}$, where $L_{\rm int}$ is the intrinsic luminosity of the cooling, non-tidally heated WD. 

Figures \ref{63temp}-\ref{69temp} display the temperature of a $0.6 M_\odot$ WD as a function of orbital period, for companion masses of $0.3 M_\odot$ and $0.9 M_\odot$. For the cases using the non-linear breaking criterion of equation (\ref{nl1}) and critical layer absorption with $t_{\rm coup} = 1 {\rm yr}$, the results are very similar to those of the $0.3 M_\odot$ He WD models. When the luminosity is dominated by tidal heating, the $0.6 M_\odot$ WD models have slightly larger surface temperatures than the $0.3 M_\odot$ models. Although the value of $\dot{E}_{\rm heat}$ is comparable between the $0.3 M_\odot$ and $0.6 M_\odot$ models, the smaller radius of the $0.6 M_\odot$ WD models requires a larger surface temperature to radiate the same amount of energy. Finally, the luminosities and temperatures are usually larger for the $0.9 M_\odot$ companion, as expected from the scaling of equation (\ref{Eheat2}). 

However, for the non-linear breaking criterion of equation (\ref{nl2}) and the critical layer absorption with $t_{\rm coup} = 10^3 {\rm yr}$, the results are markedly different. For these criteria, most of the tidal heat is deposited deeper in the WD where it cannot quickly diffuse outward to be radiated away. These criteria thus lead to generally lower WD surface temperatures. However, because the tidal heat is not quickly radiated away, the layers in which the heat is deposited may heat up substantially.

The layers just above the He-H composition gradient are primarily composed of degenerate hydrogen. If these layers are able to trap enough heat, their temperature will increase until hydrogen fusion begins. Due to the degeneracy of the hydrogen, the ignition of fusion starts a thermonuclear runaway similar to a classical nova. Our evolutions show that these tidally induced novae occur in our two cooler $0.6 M_\odot$ CO WD models and in our coolest $0.3 M_\odot$ He WD model. The novae occur only if the gravity waves deposit their heat near the base of the hydrogen layer, i.e., only for the heat deposition criteria which yield $10^{-5} M_\odot \lesssim \Delta M_B \lesssim 10^{-3} M_\odot$ for our CO WD model and $10^{-4} M_\odot \lesssim \Delta M_B \lesssim 10^{-2} M_\odot$ for our He WD model. Novae do not occur for the warmer models because the hydrogen is not degenerate, so these models are able to burn the hydrogen stably. 

In a thermonuclear runaway event, most of the hydrogen will be burned to helium or will be ejected from the system (Truran 2002). We do not attempt to predict a detailed observational signal of such an event, other than to speculate that it will be very similar to a classical nova. The thermonuclear runaway may dramatically change the dynamics of subsequent tidal heat deposition. Assuming the hydrogen shell is much thinner after the nova event, tidal heat may be deposited closer to the surface where it can quickly diffuse outwards, similar to the results for our warm He WDs.

\subsection{Heating and Cooling Time Scales}
\label{time}

The effect of tidal heating can be better understood by examining the relative time scales of the WD inspiral, heating, and cooling processes. The WD inspiral time due to gravitational radiation, $t_{\rm GW}$, is given by equation (\ref{tgw}). The inspiral time scale ranges from more than a Hubble time at large orbital periods (several hours), to less than a million years at short orbital periods (less than about fifteen minutes). Any process that acts on a time scale longer than the inspiral time is irrelevant, because the WDs will have merged (or begun stable mass transfer) by the time the process makes a substantial change.

\begin{figure*}
\begin{centering}
\includegraphics[scale=.55]{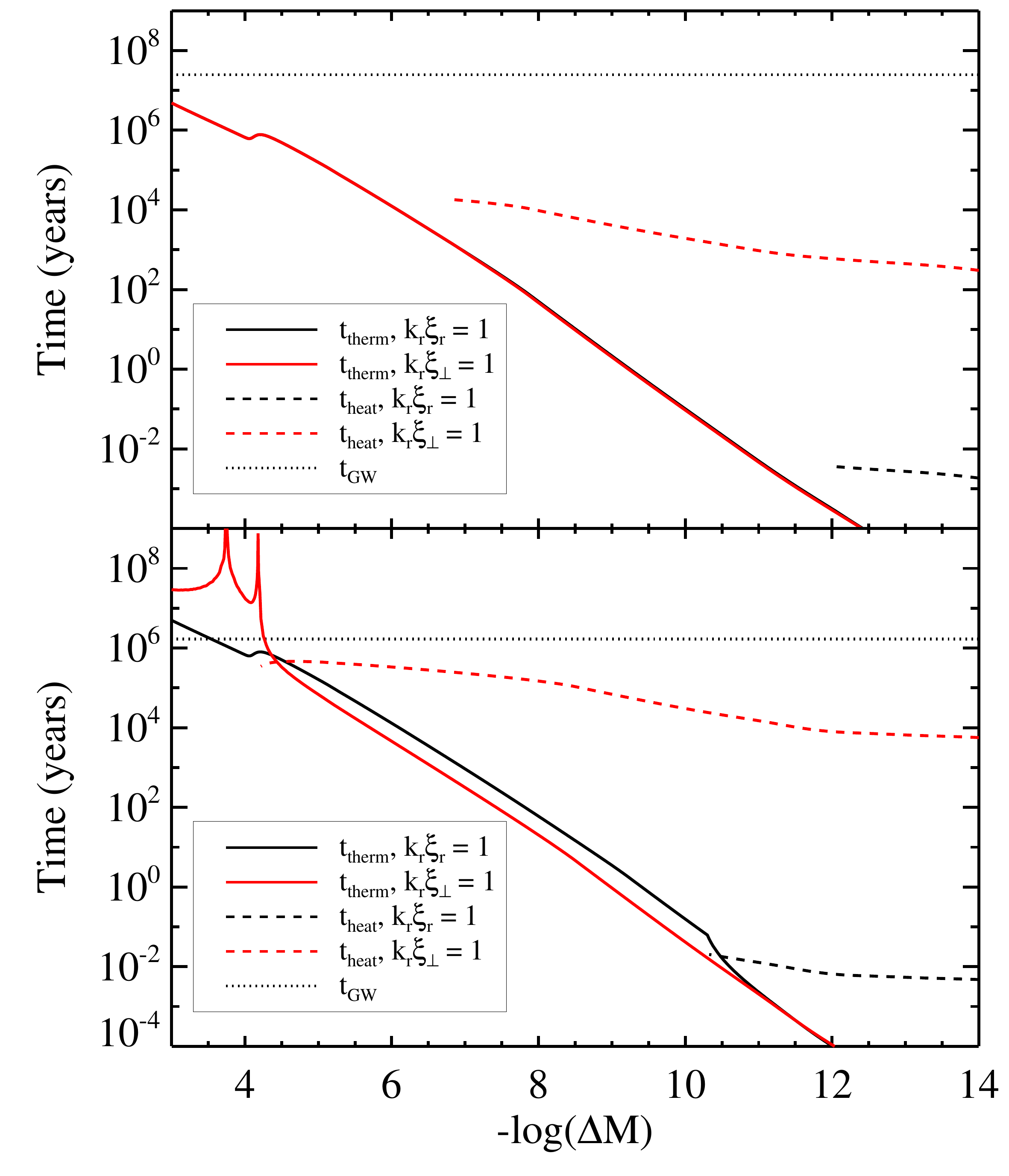}
\caption{\label{63time} The thermal time [solid lines, equation (\ref{ttherm})], heating time [dashed lines, equation (\ref{theat})], and inspiral time [dotted lines, equation (\ref{tgw})] as a function of exterior mass, $\Delta M$, in our $0.6 M_\odot$ CO WD model with initial temperature $T_{\rm eff} =10000$K. The black lines are calculated using the breaking criterion of equation (\ref{nl1}), while the red lines are calculated using the breaking criterion of equation (\ref{nl2}). The top panel is calculated for an orbital period of 45 minutes, while the bottom is for an orbital period of 15 minutes. Both panels are calculated for a $0.3 M_\odot$ companion.}
\end{centering}
\end{figure*}

The temperature evolution of the WD is determined not only by the amount of tidal heat deposited, but also by the rate at which that heat is able to diffuse to the surface. The time scale on which a shell of material heats up (in the absence of cooling) is 
\be 
\label{theat}
t_{\rm heat}(r,\Omega) = \frac{c_p T}{\dot{\varepsilon}_{\rm heat}},
\ee
where $c_p$ is the specific heat at constant pressure. The radial profile of the heating time depends primarily on the magnitude and radial dependence of the tidal heating. As tidal heat is deposited in a shell of the WD, it will diffuse on a thermal time scale,
\be 
\label{ttherm}
t_{\rm therm}(r,\Omega) = \frac{P c_p T}{g F},
\ee
where $P$ is the pressure and $F$ is the heat flux through the shell. The thermal time scale has a very sensitive dependence on the depth of the shell in question (see Figure \ref{63time}). In the core of the WD, $t_{\rm therm} \approx 10^{9} {\rm years}$, whereas near the surface, $t_{\rm therm} \ll 1 \ {\rm year}$.   

Figure \ref{63time} shows a plot of $t_{\rm GW}$, $t_{\rm heat}$, and $t_{\rm therm}$ as a function of $\Delta M$ for our $0.6 M_\odot$ WD with a $0.3 M_\odot$ companion at orbital periods of 45 and 15 minutes. At long orbital periods, $t_{\rm therm} \ll t_{\rm heat}$ at all radii. The tidal heat is able to quickly diffuse to the surface and be radiated away. The temperature of the WD reaches a thermal equilibrium such that tidal heat is radiated at the same rate it is deposited. We find this is also the case for our $0.3 M_\odot$ He WD model at most orbital periods. 

At short orbital periods, $t_{\rm therm} \approx t_{\rm heat}$ near the base of the heat deposition zone. The temperature profile will adjust so as to re-establish thermal equilibrium such that $t_{\rm therm} < t_{\rm heat}$, thereby changing the internal structure of the star. When $\Delta M_B$ is calculated with equation (\ref{nl1}), the star is able to adjust to the heating by steepening its temperature gradient (thus increasing its luminosity) such that $t_{\rm therm} \lesssim t_{\rm heat}$ at all radii.\footnote{In Figure \ref{63time}, the value of $t_{\rm therm}$ goes to infinity at some values of $\Delta M$ because the heat flux goes to zero at these locations. This can occur when the amount of tidal heat diffusing inwards from the surface is equal to the amount of intrinsic WD heat diffusing outwards from the core (i.e., there is a local temperature minimum). It can also occur where the tidal heat diffuses equally in both directions (i.e., there is a local temperature maximum).} However, when $\Delta M_B$ is calculated with equation (\ref{nl2}), the star is unable to reach thermal equilibrium and $t_{\rm therm} > t_{\rm heat}$ at the base of the heat deposition zone. The shell at $\Delta M \approx 10^{-4} M_\odot$ heats up, and eventually it reaches temperatures high enough to trigger run-away hydrogen fusion.

In principle, tidal heating may change the structure of the WD enough to alter the dynamics of wave excitation and wave breaking. However, we find that this is not the case, with the exception of the instance of a run-away hydrogen fusion event. At and below the location of wave excitation ($\Delta M \gtrsim 6 \times 10^{-3}$), $t_{\rm GW} \ll t_{\rm heat}$, so these layers are not affected by tidal heating. At shallower depths ($\Delta M \lesssim 6 \times 10^{-5} M_\odot$), significant heating may occur, creating large temperature gradients. However, in our evolutions no interior convection zone forms, despite the large temperature gradients. We thus conclude that the locations of wave excitation and breaking will not be significantly altered by the tidal heating, except in the case of a run-away hydrogen fusion event.

\section{Discussion}
\label{discussion}

\begin{figure*}
\begin{centering}
\includegraphics[scale=.55]{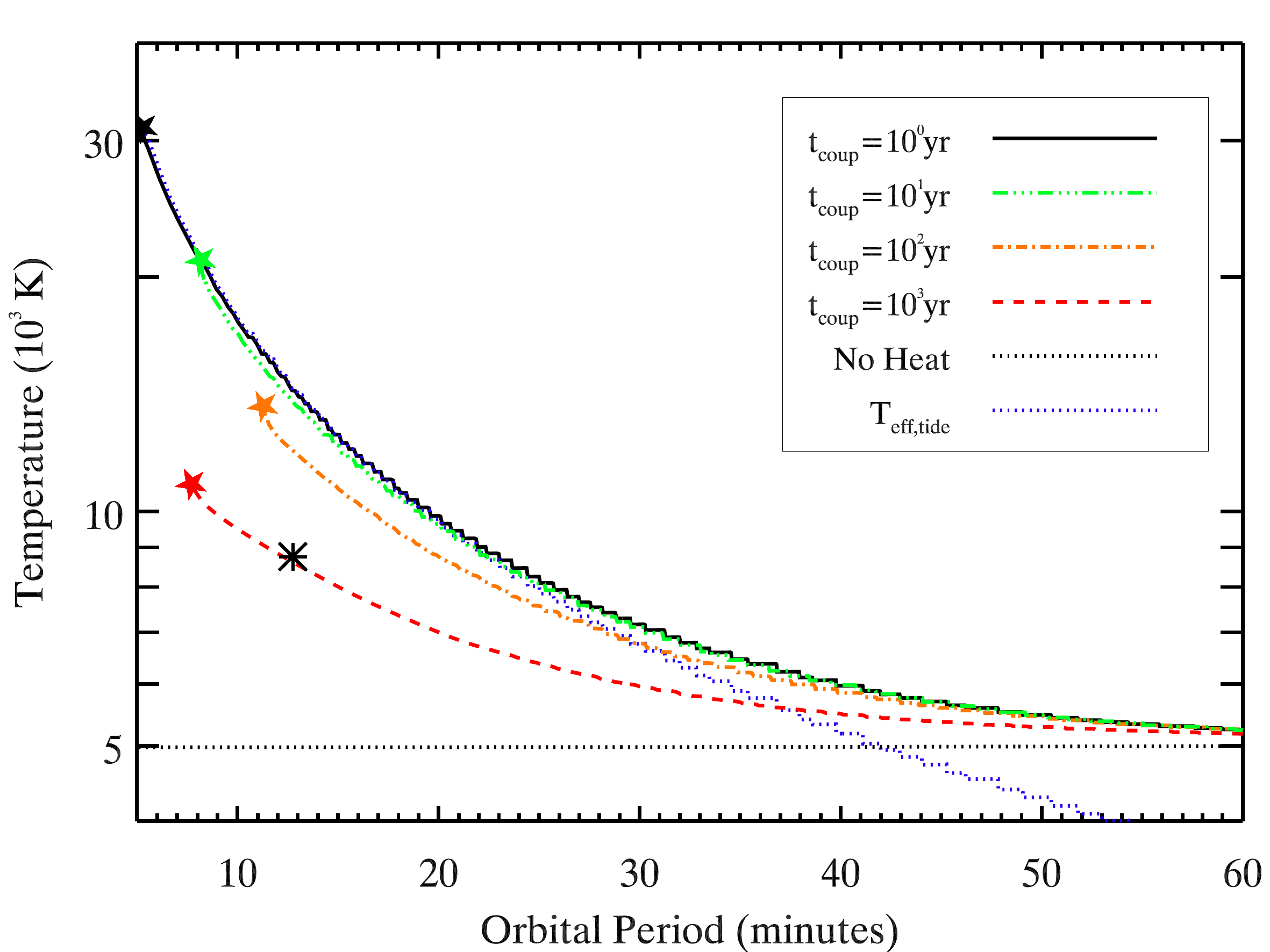}
\caption{\label{63tempT5000} Same as the top right panel of Figure \ref{63temp}, except that we have also included the temperature calculated using $t_{\rm coup} = 10^2$yr (orange dot-dashed line) and $t_{\rm coup} =10$yr (green dot-dot-dashed line). The asterisk marks the position of the secondary in SDSS J0651+2844.}
\end{centering}
\end{figure*}

We have calculated the amplitude of tidally excited gravity waves in a $0.3 M_\odot$ He WD in a compact binary system, using an outgoing radiative boundary condition. This amplitude translates directly into the rate at which tidal energy and angular momentum are transferred into the WD as a function of the tidal forcing frequency. As in the case of CO WDs (see Paper II), we find that the tidal torque and energy flux depends on the tidal forcing frequency in an erratic way. On average, the dimensionless tidal dissipation rate $F(\omega)$ [related to the inverse of the tidal quality factor, see equation (\ref{Jdot})], is several orders of magnitude smaller for He WDs than for the $0.6M_\odot$ CO WDs studied in Paper II. This difference arises from the larger entropy gradients in low-mass WDs, causing decreased coupling between gravity waves and the tidal potential [see equation (\ref{f})]. Nevertheless, since low-mass He WDs have larger radii, we find that the critical orbital frequency above which spin-orbit synchronization starts is similar in He and CO WDs, both occurring at an orbital period of about an hour. Furthermore, the amount of tidal heat deposited is similar in He and CO WDs, exceeding the intrinsic luminosity at short orbital periods ($P \lesssim 30$ minutes).

We have also estimated the location of tidal heating due to the non-linear breaking of outgoing gravity waves or resonant absorption at a critical layer in the WD envelope. The tidal heat is likely deposited in the outer layers of the WD with $\Delta M \lesssim 10^{-2} M_\odot$, although the precise location depends on the details of non-linear wave breaking and the efficiency of rotational coupling between the WD core and surface layers. We have evolved WD models to evaluate the effect of tidal heating as function of orbital period. Tidal heating is unimportant at long orbital periods ($P_{\rm orb} \gtrsim 1$ hour), but can dominate the luminosity of the WD at small orbital periods ($P_{\rm orb} \lesssim 15$ minutes). We have also found that if the tidal heat is deposited deep enough in a WD (near the He-H composition gradient), it may build up enough to trigger a nova-like hydrogen burning event (a \textquotedblleft tidal nova"). 

Our theory can be constrained by comparing the predictions of our tidal heating calculations to observed compact WD binaries. The two 39 minute systems J0106-1000 (Kilic et al. 2011a) and J1630+4233 (Kilic et al. 2011b), and the 12.75 minute system J0651+2844 (Brown et al. 2011) provide the best opportunities. We find that the warm temperatures of the He WD primaries in the two 39 minute systems ($T_{\rm eff}=16490$K in J0106 and $T_{\rm eff}=14670$K in J1630) can not be explained by tidal heating. These WDs are likely young or kept warm by residual hydrogen shell burning (Steinfadt et al. 2010b). The unseen companions in these two systems do not have measured effective temperatures, but they are unlikely to be substantially increased by tidal heating.

The 12.75 minute system J0651 likely exhibits strong tidal heating signatures. This system is composed of a primary He WD with $T_{\rm eff}=16530$K and $M=0.26 M_\odot$, and a secondary CO WD with $T_{\rm eff}\approx 8700$K and $M=0.50 M_\odot$ (Hermes et al. 2012). Comparison with Figure \ref{36temp} indicates that the luminosity of the primary is mostly due to intrinsic heat, but that tidal heat may slightly increase its temperature and luminosity. In contrast, comparison with Figure \ref{63tempT5000} indicates the luminosity of the secondary is likely dominated by tidal heat. In fact, Figure \ref{63tempT5000} predicts that $T_{\rm eff}$ is typically {\it larger} than the observed surface temperature, except for relatively deep tidal heat deposition. 

Our results indicate that the secondary in J0651 is fairly old (it takes a WD several $10^9$yr to cool to $T_{\rm eff} =5000$K), and that tidal heat deposition occurs relatively deep in the star, at $\Delta M_B \gtrsim 10^{-4} M_\odot$. Such deep tidal heat deposition could arise due to a thick critical layer created by inefficient rotational coupling between the surface layer and the core, or it could be due to non-linear wave breaking at the He-H composition gradient. Furthermore, Figure \ref{63tempT5000} indicates that a tidal nova may occur in the future of J0651.

However, there are other possible explanations for the cool observed surface temperature of the secondary in J0651. One possibility is that the value of $\dot{E}_{\rm heat}$ used in our calculations is too large. This would require the value of $\Omega_c$ to be smaller than suggested by our calculations, i.e., tidal effects become important at {\it longer} orbital periods than we have predicted. Stronger tidal effects may be possible if some other tidal dissipation mechanism (e.g., inertial waves, non-linear wave excitation, or spin-up via resonance locking, see Burkart et al. 2012 and discussion below) plays an important role in tidal synchronization. This would cause the tidal heating rate at shorter periods to become {\it smaller}, which could explain the low observed value of $T_{\rm eff}$.

We find that the ignition of run-away hydrogen fusion is a general characteristic of the tidal heating process for CO WDs with hydrogen envelopes, provided that a significant fraction of the tidal heat is deposited near the base of the hydrogen shell. Thermonuclear runaway only occurs in WDs with initial surface temperatures $T_{\rm eff} \lesssim 1.2 \times 10^4$K, otherwise tidal heating promotes steady hydrogen burning. The thermonuclear runway usually occurs at orbital periods $5 {\rm min} \lesssim P_{\rm orb} \lesssim 20 {\rm min}$, depending on the location of heat deposition, initial temperature of the WD, and companion mass. We speculate that the thermonuclear runaway will create an event very similar to a classical nova. Since most of the hydrogen would likely be burned or ejected in such an event, recurrent novae are unlikely. Thus, the occurrence rate of these tidally induced novae may be comparable to that of WD mergers involving a CO WD. 

Finally, we note that we have ignored the effect of mixing in our WD evolutions. Substantial mixing may be caused by the turbulence of breaking gravity waves or by shear instabilities due to differential rotation. If mixing occurs faster than the gravitational settling time, the composition gradients inside the WD may be smoothed out. Since the composition gradients play important roles in the excitation and breaking of gravity waves, substantial mixing may change the dynamics of the tidal synchronization and heating process. We have also ignored the effect of crystallization on the wave dynamics. Although this issue is unlikely to affect He WDs because of their long cooling times, it may be important for cool CO WDs.

Soon after our paper was submitted, Burkart et al. (2012) posted an article on arXiv studying tides in WD binaries. In addition to examining gravity waves that break [using the criterion of equation (\ref{nl1})] in systems with short orbital periods, they discussed resonance locking with WD g-modes at larger orbital periods $(P \gtrsim 1 {\rm hr})$. This may cause substantial tidal spin-up of the WD in the orbital period range $1 {\rm hr} \lesssim P \lesssim 4 {\rm hr}$ if the associated gravity waves do not break. At shorter orbital periods $(P \lesssim 45 {\rm min})$, Burkart et al. (2012) found that gravity waves indeed break, producing tidal torques and spin evolution similar to ours. Note that if gravity waves break via the non-linearity criterion of equation (\ref{nl2}), then the resonance locking regime will be limited to long orbital periods. This will reduce the effectiveness of tidal spin-up prior to the onset of continual gravity wave breaking, bringing their results more in line with ours for $P\sim{\rm hrs}$.  In another recent paper on WD binaries, Valsecchi et al. (2012) discussed possible \textquotedblleft anti-resonance locking" of g-modes in the linear theory.  The nature of this \textquotedblleft anti-resonance" locking is not clear, nor is the parameter regime where such locking is effective. Overall, it appears that there is a general agreement that at short orbital periods ($P \lesssim 1 {\rm hr}$), the WD spins up on the orbital decay timescale, with $\Omega_s/\Omega$ approaching unity as $\Omega$ increases, but maintaining an approximately constant $(\Omega-\Omega_s)$ until binary merger or the onset of mass transfer [see Figure 4 or equation (\ref{eq:omegas}); see also Section 8.1 of Paper II]. 

Future observations and simulations can further constrain these theories. We are hopeful that more compact WD systems with $P_{\rm orb} \lesssim 30$ minutes will be discovered in the near future. Measurements of the masses, luminosities, temperatures, and spins of these systems will provide more data points for comparison with our theory. Finally, simulations of gravity waves propagating through a WD envelope could be used to understand the non-linear criterion that governs wave breaking in WDs and could more conclusively determine where tidal heat is deposited in a WD.

\section*{Acknowledgments}

We thank Bill Paxton, Lars Bildsten, and Eliot Quataert for useful discussion. JF acknowledges the hospitality (Fall 2011)
of the Kavli Institute for Theoretical Physics at UCSB (funded by the NSF through Grant 11-Astro11F-0016) where part of the work was carried out. This work has been supported in part by NSF grants AST-1008245 and AST-1211061, and NASA grants NNX12AF85G, NNX10AP19G, and NNX11AL13H.

\appendix

\section{Calculation of Equilibrium Tide}

In this appendix we present our improved method for isolating the dynamical and equilibrium components of the tidal perturbation. Using the Cowling approximation, the oscillation equations are 
\be
\label{xir'}
\frac{1}{r^2} \big(r^2 \xi_r \big)' - \frac{g}{c_s^2} \xi_r + \frac{1}{\rho c_s^2} \bigg(1 - \frac{L_l^2}{\omega^2} \bigg) \delta P - \frac{l(l+1)U}{\omega^2 r^2} = 0,
\ee
and
\be
\label{p'}
\delta P' + \frac{g}{c_s^2}\delta P + \rho \big(N^2 - \omega^2\big)\xi_r + \rho U' = 0,
\ee
where $U$ is the tidal potential produced by the companion, the $'$ denotes $d/dr$, and $g$ is the gravitational acceleration. The other perturbation variables are related to $\delta P$ and $\xi_r$ by 
\be
\label{xiperp}
\xi_\perp = \frac{1}{r \omega^2} \bigg( \frac{\delta P}{\rho} + U \bigg),
\ee
\be
\delta \rho = \frac{1}{c_s^2} \delta P + \frac{\rho N^2}{g} \xi_r.
\ee

Using equation (\ref{xiperp}), equations (\ref{xir'}) and (\ref{p'}) may be rewritten
\be
\label{xir'2}
\frac{1}{r^2} \big(r^2 \xi_r \big)' - \frac{g}{c_s^2} \xi_r + \bigg(\frac{r\omega^2}{c_s^2} - \frac{l(l+1)}{r} \bigg) \xi_\perp - \frac{U}{c_s^2} = 0,
\ee
\be
\label{xit'}
\xi_\perp ' + \bigg(\frac{1}{r} - \frac{N^2}{g}\bigg) \xi_\perp + \frac{N^2 - \omega^2}{r \omega^2} \xi_r + \frac{N^2}{r g \omega^2} U = 0.
\ee

The zeroth order solution to the equilibrium tide can be found by taking the limit $\omega=0$ in equations (\ref{xir'2}) and (\ref{xit'}), yielding (see also Goldreich \& Nicholson 1979)
\be
\xi_r^{\rm eq,0} = -U/g
\ee
and
\be
\xi_\perp^{\rm eq,0} = -(Ur^2/g)'/[l(l+1)r].
\ee
Defining $\xi_r = \xi_r^{\rm eq,0} + \bar{\xi}_r$ and likewise for $\xi_\perp$, we substitute into equations (\ref{xir'2}) and (\ref{xit'}) to find
\be
\label{xird'}
\frac{1}{r^2} \big(r^2 \bar{\xi}_r \big)' - \frac{g}{c_s^2} \bar{\xi}_r +  \bigg(\frac{r \omega^2}{c_s^2} - \frac{l(l+1)}{r} \bigg) \bar{\xi}_\perp + \frac{r \omega^2}{c_s^2}\xi_\perp^{\rm eq,0} = 0,
\ee
and
\be
\label{xitd'}
\bar{\xi}_\perp ' + \bigg(\frac{1}{r} - \frac{N^2}{g}\bigg)\bar{\xi}_\perp + \frac{N^2-\omega^2}{r \omega^2} \bar{\xi}_r + \bigg[ \xi_\perp^{\rm eq,0'} + \bigg(\frac{1}{r} - \frac{N^2}{g}\bigg) \xi_\perp^{\rm eq,0} - \frac{1}{r} \xi_r^{\rm eq,0} \bigg].
\ee
To find the first order term of the equilibrium tide, we again take the limit of $\omega=0$ in equations (\ref{xird'}-\ref{xitd'}) to find
\be
\label{xireq1}
\xi_r^{\rm eq,1} = \frac{r \omega^2}{N^2 - \omega^2} \bigg[\bigg(\frac{N^2}{g} - \frac{1}{r}\bigg) \xi_\perp^{\rm eq,0} + \frac{1}{r} \xi_r^{\rm eq,0} - \xi_\perp^{\rm eq,0'} \bigg],
\ee
and
\be
\label{xiperpeq1}
\xi_\perp^{\rm eq,1} = \frac{1}{l(l+1)} \bigg[\bigg(2 - \frac{gr}{c_s^2} \bigg) \xi_r^{\rm eq,1} + r \xi_r^{\rm eq,1'} + \frac{r^2\omega^2}{c_s^2} \xi_\perp^{\rm eq,0} \bigg].
\ee

We find that higher order terms are not essential for our purposes. Moreover, these terms typically involve multiple derivatives of stellar properties that are difficult to compute from a grid of stellar quantities. The total equilibrium tide is $\bxi^{\rm eq} = \bxi^{\rm eq,0} + \bxi^{\rm eq,1} +  \mathcal{O}(\omega^4)$. At our outer boundary, the dynamical tide is computed using $\bxi^{\rm dyn} = \bxi - \bxi^{\rm eq}$.

\section{General Scaling for Spin-Orbit Synchronization and Tidal Heating}
\label{genscale}

In Section 2.3, we showed that the averaged dimensionless tidal torque on a He WD scales with the tidal forcing frequency as $F\propto \omega^6$. For a CO WD, the scaling is $F\propto\omega^5$ (see Paper II). In general, we may parameterize the tidal torque by 
\be
\label{Ttide}
T_{\rm tide}=\dot J_z= T_0 \hat{f} \hat{\omega}^n
\ee
with the dimensionless values $\hat{f}$ and $n$ determined from the dynamics of gravity wave excitation and dissipation within the WD.  Using equation (\ref{Ttide}), we can examine the general behaviors of spin-orbit synchronization and tidal heating in a compact binary undergoing gravitational radiation-driven orbital decay (see also Section 8.1 of Paper II). 

Combining equations (\ref{Jdot}) and (\ref{Omegadot}) as we did in Section \ref{rotation}, we find that tides begin to synchronize the star at a critical orbital frequency
\be
\label{omegacgen}
\Omega_c = \bigg[ \frac{2.4}{2^n \sqrt{2}} \frac{\kappa}{\hat{f}} \bigg( \frac{R_s}{R}\bigg)^{5/2} q_M \bigg]^{\frac{3}{3n+1}} \Omega_{\rm dyn}.
\ee
Here, $R_s = 2GM/c^2$, $q_M = (M+M')^{5/3}/(M^{2/3}M')$, and $\Omega_{\rm dyn} = \sqrt{GM/R^3}$. Thus, less compact stars (with larger $R/R_s$) begin synchronization at larger orbital periods. Note that if $n \go 3$ as for WDs, the critical frequency $\Omega_c$ does not depend sensitively on $\hat{f}$ or $\hat{\kappa}$, and only a very rough estimate of these quantities is necessary.

The tidal heating rate is given by equation (\ref{Eheat}). For $\Omega > \Omega_c$ (see Paper II),
\be
\Omega_s\simeq \Omega-\Omega_c \left({\Omega_c\over\Omega}\right)^{1/(3n)},
\label{eq:omegas}
\ee
and combining equations (\ref{Eheat}), (\ref{omegacgen}), and (\ref{eq:omegas}) yields
\be
\label{eheatgen}
\dot{E}_{\rm heat} = \bigg[ \frac{2.4}{2^n \sqrt{2}} \frac{\kappa^{n+1}}{\hat{f}} \bigg( \frac{R_s}{R}\bigg)^{5/2} q_{M2} \bigg]^{\frac{1}{n}} \bigg(\frac{\Omega}{\Omega_{\rm dyn}}\bigg)^{\frac{n-1}{3n}} \dot{E}_{\rm GW},
\ee
where $q_{M2} = (M+M')^{(5+n)/3} M^{2(n-1)/3}/M'^{n+1}$. Note that for $n\go3$ as it is for WDs, the tidal heating rate is very insensitive to the value of $\hat{f}$. At typical orbital periods ($\Omega \ll \Omega_{\rm dyn}$), the tidal heating rate is a small fraction of the energy loss rate due to gravitational waves (especially for non-compact stars), but it becomes more significant at shorter orbital periods.

\section{Estimating the Core-Envelope Coupling Time Scale}
\label{tcoup}

In the two zone model (see Section \ref{critical}), the thickness of the synchronized envelope is dependent on the parameter $t_{\rm coup}$. In stably stratified stars like WDs, angular momentum can be transported by magnetic fields. In the presence
of a poloidal field $B$ connecting the core and envelope, $t_{\rm coup}$ can be estimated from the Alfven wave crossing time,
\be
\label{ta}
t_A = \int^R_0 \!dr \frac{\sqrt{4\pi\rho}}{B}.
\ee
We find $t_A \approx 10^2 \ {\rm yr} (10^3 {\rm G}/B)$ for our CO WD model, and $t_A \approx 50 \ {\rm yr} (10^3 {\rm G}/B)$ for our He WD model. 

For WDs without an intrinsic magnetic field, angular momentum may be transported via the Tayler-Spruit dynamo (Spruit 2002). To estimate $t_{\rm coup}$, we calculate the effective viscosity $\nu_{TS}$ for angular momentum transport using the the method outlined in Spruit (2002). For simplicity, we calculate $\nu_{TS}$ without including the effects of composition gradients in the WD to find
\be
\label{nuts}
\nu_{TS} = r^2 \Omega_s \bigg(\frac{\Omega_s}{N}\bigg)^{1/2} \bigg(\frac{\kappa_T}{r^2 N}\bigg)^{1/2},
\ee
where $\kappa_T$ is the heat diffusivity. This viscosity does not depend on the local shear, and is thus independent of the precise rotation profile of the star.\footnote{The Tayler-Spruit dynamo requires a minimum local shear to operate (see Spruit 2002), causing significant shear to persist if the dynamo dominates angular momentum transport. This shear may significantly impact wave dynamics, in addition to causing viscous heating and mixing.}

We then estimate the coupling time from the Tayler-Spruit dynamo to be
\be
\label{tts}
t_{TS} = \int^R_0 \!dr \frac{R-r}{\nu_{TS}}.
\ee
We find $t_{TS} \approx 2\times10^{3}\,{\rm yr}\,(P/30{\rm min})^{3/2}$ for both $0.3 M_\odot$ He and $0.6 M_\odot$ CO WDs. The values of $t_A$ and $t_{TS}$ convince us that $t_{\rm coup} \lesssim 10^{3}\,{\rm yr}$ for the short orbital periods of interest.

\def\apj{{Astrophys. J.}}
\def\apjs{{Astrophys. J. Supp.}}
\def\mnras{{Mon. Not. R. Astr. Soc.}}
\def\prl{{Phys. Rev. Lett.}}
\def\prd{{Phys. Rev. D}}
\def\apjl{{Astrophys. J. Let.}}
\def\pasp{{Publ. Astr. Soc. Pacific}}
\def\aapr{{Astr. Astr. Rev.}}


\end{document}